\documentclass[11pt,a4paper]{article}

\usepackage{Setting/jhepCAI190929}
\usepackage{ifpdf}

\usepackage{afterpage}
\usepackage[utf8]{inputenc}

\usepackage[T1]{fontenc} % if needed

\usepackage{latexsym}
\usepackage{amsmath}
\usepackage{graphicx}
\usepackage{subfigure}
\usepackage{dcolumn}
\usepackage{bm}
\usepackage{amssymb}
\usepackage{latexsym}
\usepackage{datetime}
\usepackage{scrtime}
\usepackage{color}
\usepackage[dvipsnames, svgnames, x11names]{xcolor}

\def\be{\begin{equation}}
\def\ee{\end{equation}}
\def\ba{\begin{eqnarray}}
\def\ea{\end{eqnarray}}

\def\nn{\nonumber}
\def\lf{\left}
\def\rt{\right}

\renewcommand{\(}{\left(}
\renewcommand{\)}{\right)}
\renewcommand{\[}{\left[}
\renewcommand{\]}{\right]}

\begin{document}
%\pagestyle{myplain}	
%\begin{flushleft}
%	{\footnotesize
%		UCAS/2019/01
%	}
%\end{flushleft}

\title{Generating enhanced primordial GWs during inflation with intermittent violation of NEC and diminishment of GW propagating speed}

\author{Yong Cai,$^{1}$ }
\author{Yun-Song Piao$^{2,3,4,5}$}

\affiliation{$^1$ School of Physics and Microelectronics, Zhengzhou University, Zhengzhou, Henan 450001, China}

\affiliation{$^2$ School of Physics, University of Chinese Academy
	of Sciences, Beijing 100049, China}

\affiliation{$^3$ School of Fundamental Physics and Mathematical
	Sciences, Hangzhou Institute for Advanced Study, UCAS, Hangzhou
	310024, China}

\affiliation{$^4$ International Center for Theoretical Physics
	Asia-Pacific, Beijing/Hangzhou, China}

\affiliation{$^5$ Institute of Theoretical Physics, Chinese
	Academy of Sciences, P.O. Box 2735, Beijing 100190, China}

\emailAdd{yongcai\_phy@outlook.com}
\emailAdd{yspiao@ucas.ac.cn}

\abstract{
We investigate both the null energy condition (NEC) violating scenario and the $c_T$-diminishing scenario for generating enhanced power spectrum of primordial gravitational waves (GWs) during inflation, where $c_T$ is the propagating speed of primordial GWs. Both of these two scenarios can be realized stably with theories beyond Horndeski, hence can be uniformly implemented within the framework of the effective field theory. We calculate the power spectrum of primordial GWs by assuming that the inflationary Universe undergoes three phases, where the violation of NEC or the diminishment of $c_T$ occurs in the intermediate phase. A template of the spectrum is given for the NEC-violating scenario. We also discuss the underlying relation and discrepancy between these two scenarios with a disformal transformation.
}

%\keywords{Keyword1,Keyword2}

\maketitle

%\tableofcontents

\section{Introduction}

The direct detections of gravitational waves (GWs) from binary black holes \cite{LIGOScientific:2016aoc} and a binary neutron star inspiral \cite{LIGOScientific:2017vwq} by the LIGO and Virgo collaborations have opened a new window for exploring our Universe and the theory of gravity.
Recently, evidence for a common-spectrum process, which might be interpreted as a stochastic GW background (see e.g. \cite{Ellis:2020ena,Li:2020cjj,Addazi:2020zcj,Tahara:2020fmn,Kuroyanagi:2020sfw,Brandenburg:2021tmp,Yi:2021lxc,Lewicki:2021xku,Bian:2020urb,Sun:2021yra,Vagnozzi:2020gtf,Samanta:2020cdk} for related studies, see also \cite{Chen:2019xse,Chen:2021wdo,Wu:2021kmd}), was reported by the North American Nanohertz Observatory for Gravitational Waves (NANOGrav) collaboration in their 12.5-year dataset \cite{NANOGrav:2020bcs} and also reported later by the Parkes Pulsar Timing Array (PPTA) collaboration \cite{Goncharov:2021oub}.
A wide multi-frequency range of observations aimed at searching for GW signals will lead us to the era of GW astronomy.

The primordial GWs \cite{Starobinsky:1979ty,Rubakov:1982df} predicted by inflation have yet to be confirmed by future observations. In the slow-roll inflation scenario, the power spectrum of primordial GWs is usually assumed as a power-law. As a result, the primordial GWs can hardly be detected by the pulsar timing arrays (PTA) or laser interferometers in the near future, since the tensor-to-scalar ratio $r$ is severely constrained by the data of cosmic microwave background (CMB) B-mode polarization \cite{Planck:2018vyg}. However, the primordial GW background spans a broad frequency-band ($10^{-18}-10^{10}\,{\rm Hz}$), the structure of the power spectrum and the underlying physics might be far richer than expected \cite{Cai:2020qpu,Cai:2021yvq,Benetti:2021uea}.

%During inflation, there exists various mechanisms that are able to generate primordial GWs with enhanced power spectrum at frequency higher than that of the CMB window.

%In this paper, we will focus on violation of the null energy condition (NEC) \cite{Piao:2003ty,Piao:2004tq,Baldi:2005gk,Piao:2006jz,Li:2016awk,Cai:2020qpu} and diminishment of propagating speed of primordial GWs \cite{Cai:2015dta,Cai:2015ipa,Cai:2015yza,Cai:2016ldn} in an inflationary background.

Primordial null energy condition (NEC) violation (i.e., $T_{\mu\nu}k^\mu k^\nu<0$, where $T_{\mu\nu}$ is the energy-momentum tensor and $k^\mu$ is any null vector) could occur during the inflationary era \cite{Piao:2003ty,Piao:2004tq,Baldi:2005gk,Piao:2006jz,Li:2016awk} (see also \cite{Creminelli:2010ba,Liu:2011ns,Wang:2012bq,Liu:2012ww,Creminelli:2012my,Hinterbichler:2012fr,Hinterbichler:2012yn,Liu:2014tda,Pirtskhalava:2014esa,Nishi:2015pta,Kobayashi:2015gga,Cai:2016gjd,Nishi:2016ljg,Ageeva:2018lko,Mironov:2019qjt,Ageeva:2020gti,Ilyas:2020zcb,Zhu:2021ggm} for the Genesis scenario) and generate GW background with rich features in the power spectrum, see e.g. \cite{Cai:2020qpu} for a Great Wall–like spectrum.
The GW power spectrum can be enhanced greatly at frequencies higher than the CMB window since the NEC violation indicates $\dot{H}>0$ for a homogeneous and isotropic cosmological background. Fully stable NEC violation can be realized in scalar-tensor theories beyond Horndeski, as discovered in the exploration of the effective field theory (EFT) of nonsingular cosmology \cite{Cai:2016thi,Creminelli:2016zwa,Cai:2017tku,Cai:2017dyi}.

Additionally, the propagating speed of GWs (i.e., $c_T$) can be time-dependent in some gravity theories beyong general relativity. Although the joint detection of GWs and the gamma rays from the binary neutron star merger GW170817 has strongly supported that the propagating speed of GWs is $c_T=1$ at present \cite{LIGOScientific:2018dkp,Creminelli:2017sry,Sakstein:2017xjx,Baker:2017hug,Langlois:2017dyl}, with deviations smaller than ${\cal O}(10^{-15})$, it still cannot place stringent bounds on $c_T$ in the very early Universe, see e.g., \cite{deRham:2018red,deRham:2019ctd,deRham:2020zyh,Tian:2019vkc}. The nontrivial variation of $c_T$ is able to imprint interesting features in the power spectrum of GWs \cite{Cai:2015dta,Cai:2015ipa}, see also \cite{Giare:2020vss,Bernal:2020ywq,Cai:2020ovp}.
An enhanced (or blue) power spectrum of primordial GWs may also be attributed to the diminishment of $c_T$ \cite{Cai:2015yza,Cai:2016ldn,Giovannini:2018zbf,Giovannini:2018nkt,Mishima:2019vlh,Capurri:2020qgz,Giovannini:2021uvh}, which can be implemented with scalar-tensor theories beyond Horndeski. Intriguingly, it is found that a scenario in which $c_T$ gradually diminishes during inflation is actually a disformal dual to the superinflation \cite{Cai:2016ldn}.

In this paper, we investigate both the NEC-violating scenario and the $c_T$-diminishing scenario for generating enhanced power spectrum of primordial GWs during inflation within the uniform framework of EFT. In Sec. \ref{sec:EFTm}, we briefly review the EFT method.
In Sec. \ref{Sec:NECV01}, we explicitly calculate the power spectrum of primordial GWs for a scenario in which the NEC-preserving slow-roll inflation is followed by an NEC-violating expanding phase and a later NEC-preserving slow-roll inflation with a larger Hubble parameter, while $c_T=1$ throughout. Templates of such spectra are provided.
We also calculate the power spectrum of primordial GWs in Sec. \ref{sec:cTdim} for a scenario in which $c_T$ decreases with time during an intermediate phase of the slow-roll inflation, while the NEC is preserving throughout. The underlying relation and discrepancy between these two scenarios are discussed with a disformal transformation in Sec. \ref{Sec:relation01}.
The speed of light is set as $c=1$ throughout this paper.

%By assuming that the inflationary Universe undergoes three phases, where the violation of NEC or the diminishment of $c_T$ occurs in the intermediate phase, we calculate the power spectrum of primordial GWs. A template of the spectrum is given for the NEC-violating scenario. We also discuss the underlying relation and discrepancy between these two scenarios.

%We investigate two methods for generating enhanced power spectra of primordial gravitational waves (GWs) during inflation, one requires intermittent null energy condition (NEC) violations during inflation, while the other one turns to a diminishing propagating speed (i.e., $c_T$) of primordial GWs.
%Both of these two methods can be realized stably with theories beyond Horndeski, hence can be uniformly implemented within the framework of the effective field theory.
%We explicitly calculate the power spectra of primordial GWs for a scenario in which the NEC-preserving slow-roll inflation is followed by an NEC-violating expanding phase and a later NEC-preserving slow-roll inflation with a larger Hubble parameter, while $c_T=1$ throughout. Templates of such spectra are provided. We also calculate the power spectra of primordial GWs for a scenario in which $c_T$ decreases with time during an intermediate phase of the slow-roll inflation, while the NEC is preserving throughout. The underlying relation and discrepancy between these two scenarios are discussed.

\section{The effective field theory method}\label{sec:EFTm}

%In this section, we briefly review the EFT method, the power spectra and energy density spectra of primordial GWs.

%\subsection{Action of the effective field theory}

It is convenient to use the $3 + 1$ decomposed metric
\be
d s^{2}=-N^{2} d t^{2}+h_{i j}\left(d x^{i}+N^{i} d t\right)\left(d x^{j}+N^{j} d t\right)\,,
\ee
where $N$ is the lapse function, $N^i$ is the shift vector, $h_{i j}$ is the 3-dimensional metric.
In the unitary gauge, the EFT of cosmological perturbations up to quadratic order can be written as
\ba
S&=&\int
d^4x\sqrt{-g}\Big[ {M_p^2\over2} f(t)R-\Lambda(t)-c(t)g^{00}
\nn\\
&\,&\qquad\qquad\quad +{M_2^4(t)\over2}(\delta g^{00})^2-{m_3^3(t)\over2}\delta
K\delta g^{00} -m_4^2(t)\lf( \delta K^2-\delta K_{\mu\nu}\delta
K^{\mu\nu} \rt)
\nn\\
&\,&\qquad\qquad\quad  + {\tilde{m}_4^2(t)\over
	2}R^{(3)}\delta g^{00}\Big] +S_{\rm m}[g_{\mu\nu},\psi_{\rm m}]\,,
\label{action01}
\ea
where we have disregarded those higher-order spatial derivatives; $\delta g^{00}=g^{00}+1$, $\delta K_{\mu\nu}=K_{\mu\nu}-h_{\mu\nu}H$ with $H$ being the Hubble parameter, $h_{\mu \nu} = g_{\mu \nu}+n_{\mu} n_{\nu}$ is the induced metric, $n^\mu$ is the unit normal vector of the constant
time hypersurfaces, $K_{\mu\nu}$ is the extrinsic curvature, $R^{(3)}$ is the induced 3-dimensional Ricci scalar, $S_{\rm m}$ represents the action of matter sector, which is minimally coupled to the metric $g_{\mu\nu}$.

These functions $f$, $\Lambda$, $c$, $M_2^4$, $m_3^3$, $m_4^2$ and $\tilde{m}_4^2$ could be time-dependent in general so that they are able to specify different theories of gravity. Particularly,
the Horndeski theory \cite{Horndeski:1974wa,Deffayet:2011gz,Kobayashi:2011nu} corresponds to $m_4^2= {\tilde m}_4^2$, the ``beyond Horndeski'' theory \cite{Gleyzes:2014dya} requires $m_4^2\neq {\tilde m}_4^2$, while the degenerate higher-order scalar-tensor (DHOST) theories \cite{Langlois:2015cwa} require the appearances of new operators, such as $\delta K \dot{\delta g^{00}}$, $(\dot{\delta g^{00}})^2$ and $(\partial_i{\delta g^{00}})^2$ in the action (\ref{action01}), see e.g., \cite{Langlois:2017mxy,Langlois:2017mdk}.

The background evolution of the Universe is only determined by the first line of (\ref{action01}).
Particularly, for $f(t)\equiv1$, the Friedmann equations can be given as $c(t)=-M_{p}^{2} \dot{H}$ and $\Lambda(t)=M_{p}^{2}\left(\dot{H}+3 H^{2}\right)$.
With a variety of designs of those coefficient functions, the EFT action (\ref{action01}) is versatile in applications to a variety of cosmological background, including inflation \cite{Cheung:2007st}, dark energy \cite{Gubitosi:2012hu,Gleyzes:2013ooa,Piazza:2013coa} and nonsingular cosmology \cite{Cai:2016thi,Creminelli:2016zwa,Cai:2017tku}. Furthermore, the EFT method has great advantages in the study of cosmological perturbations.

%\subsection{Primordial GW spectrum}

The derivation of quadratic actions for scalar and tensor perturbations of the action  (\ref{action01}) can be found in \cite{Cai:2016thi}.
In this paper, we will focus on the tensor perturbations $\gamma_{ij}$ (i.e., primordial GWs), which is traceless and divergence-free, i.e., $\gamma_{ii}=0=\partial_i\gamma_{ij}$, where we have set
\be h_{ij}=a^2 (e^{\gamma})_{ij} \,.
\ee
Generally, the quadratic actions of $\gamma_{ij}$ can be formulated as
\be
S^{(2)}_{\gamma}={M_p^2\over8}\int d^4xa^3
Q_T\lf[ \dot{\gamma}_{ij}^2 -c_T^2{(\partial_k\gamma_{ij})^2\over
	a^2}\rt]\,, \label{tensor-action}
\ee
where $\dot{}=d/dt$, $c_T$ is the propagating speed of primordial GWs. In order to avoid the ghost and gradient instabilities, $Q_T>0$ and $c_T^2>0$ are required, respectively.
From (\ref{action01}), we have
\be Q_T= f+{2 m_4^2\over M_p^2}\,, \qquad c_T^2=f/Q_T\,,\label{eq:QTcT01}
\ee
where $f$ and $m_4^2$ are time-dependent coefficients defined in (\ref{action01}).
When $f=1$ and $m_4^2=0$, we have $Q_T=1$ and $c_T=1$, which are same as those in general relativity.

In the momentum space, we have \be
\gamma_{ij}(\tau,\mathbf{x})=\int \frac{d^3k}{(2\pi)^{3}
}e^{-i\mathbf{k}\cdot \mathbf{x}} \sum_{\lambda=+,\times}
\hat{\gamma}_{\lambda}(\tau,\mathbf{k})
\epsilon^{(\lambda)}_{ij}(\mathbf{k}), \ee where
$\hat{\gamma}_{\lambda}(\tau,\mathbf{k})=
\gamma_{\lambda}(\tau,k)a_{\lambda}(\mathbf{k})
+\gamma_{\lambda}^*(\tau,-k)a_{\lambda}^{\dag}(-\mathbf{k})$,
the polarizations
$\epsilon_{ij}^{(\lambda)}(\mathbf{k})$ satisfy
$k_{j}\epsilon_{ij}^{(\lambda)}(\mathbf{k})=0$,
$\epsilon_{ii}^{(\lambda)}(\mathbf{k})=0$,
$\epsilon_{ij}^{(\lambda)}(\mathbf{k})
\epsilon_{ij}^{*(\lambda^{\prime}) }(\mathbf{k})=\delta_{\lambda
	\lambda^{\prime} }$ and $\epsilon_{ij}^{*(\lambda)
}(\mathbf{k})=\epsilon_{ij}^{(\lambda) }(-\mathbf{k})$;
$a_{\lambda}(\mathbf{k})$ and
$a^{\dag}_{\lambda}(\mathbf{k}^{\prime})$ satisfy $[
a_{\lambda}(\mathbf{k}),a_{\lambda^{\prime}}^{\dag}(\mathbf{k}^{\prime})
]=\delta_{\lambda\lambda^{\prime}}\delta^{(3)}(\mathbf{k}-\mathbf{k}^{\prime})$.
The power spectrum of primordial GWs is
\be
P_T={k^3\over 2\pi^2}\sum_{\lambda=+,\times}|\gamma_\lambda|^2\,,\label{eq:PT01}
\ee
which is evaluated after the perturbation modes exited their horizons, i.e., $a H /\left(c_{T} k\right) \gg 1$.

The energy density spectrum of primordial GWs (see e.g. \cite{Turner:1993vb}, see also
\cite{Boyle:2005se,Zhao:2006mm,Kuroyanagi:2014nba,Liu:2015psa}) can be given by
\be
\Omega_{\rm GW}(\tau_{0})=\frac{k^{2}}{12
a_0^2H^2_0}P_{T}(k)\lf[\frac{3
\Omega_{\rm m}j_1(k\tau_0)}{k\tau_{0}}\sqrt{1.0+1.36\frac{k}{k_{\text{eq}}}
+2.50\left( \frac{k}{k_{\text{eq}}}\right) ^{2}}\rt]^2,
\label{GW0}
\ee
where $k_{\rm eq}$ is the comoving wavenumber of the perturbation
mode that entered the horizon at the matter-radiation equality,
$\Omega_{\rm m}=\rho_{\rm m}/\rho_{\rm c}$, $\rho_{\rm c}=3M_p^2H^{2}_0$ is the critical energy density.
The observation of a stochastic GW background would provide us with significant information or constraints on the primordial GWs as well as the evolution history of the very early Universe.

\section{Enhanced power spectrum of primordial GWs from intermittent NEC violations during inflation} \label{Sec:NECV01}

Primordial NEC violation would imprint a blue-tilted power spectrum of primordial GWs. A series of intermittent NEC violations during inflation would generate a spectrum of primordial GWs with far richer structures, such as a Great Wall–like spectrum \cite{Cai:2020qpu}. An enhanced power spectrum of primordial GWs at the frequency $\sim 1/{\rm yr}$ induced by NEC violation might be compatible with the recent result of NANOGrav, provided the reported common-spectrum process could be interpreted as GWs.

In this section, we analytically calculate the power spectrum of primordial GWs in a scenario proposed by \cite{Cai:2020qpu} and provide a template of such an enhanced GW spectrum. Follow \cite{Cai:2020qpu}, we assume that after an NEC-preserving slow-roll inflation with Hubble parameter $H \simeq H_{inf1}$, the Universe experienced an NEC-violating expansion, and then enters a subsequent NEC-preserving slow-roll inflation again but with a higher Hubble scale $H$ $\left(=H_{inf2} \gg H_{inf1}\right)$.

\subsection{Set up}

Instead of explicitly implementing this scenario in scalar-tensor theories, we adopt the EFT action
\ba
S&=&\int
d^4x\sqrt{-g}\Big[ {M_p^2\over2} R-\Lambda(t)-c(t)g^{00}
\nn\\
&\,&\qquad\qquad\quad +{M_2^4(t)\over2}(\delta g^{00})^2-{m_3^3(t)\over2}\delta
K\delta g^{00}  + {\tilde{m}_4^2(t)\over
	2}R^{(3)}\delta g^{00}\Big]\,,
\label{action02}
\ea
where we have disregarded the action of the matter sector (i.e., $S_{\rm m}$), which should make a negligible contribution to the evolution of the very early Universe in our scenario.
The action (\ref{action02}) corresponds to setting $f(t)=1$ and $m_4^2(t)=0$ in action (\ref{action01}).
As a result, we have
\be
Q_T=1,\qquad c_T=1 \label{QTcT001}
\ee
in Eq. (\ref{tensor-action}). Although the covariant correspondence of action (\ref{action02}) (see \cite{Cai:2017dyi,Kolevatov:2017voe}) belongs to the ``beyond Horndeski'' theory, the propagation of GWs is exactly same as that in general relativity, at least at quadratic order.

The functions $c(t)$ and $\Lambda(t)$ can be determined in a reverse manner by the background evolution with  relations $c(t)=-M_{p}^{2} \dot{H}$ and $\Lambda(t)=M_{p}^{2}\left(\dot{H}+3 H^{2}\right)$, while $M_2^4(t)$, $m_3^3(t)$ and $\tilde{m}_4^2(t)$ can be determined or constrained by the requirement that the scalar perturbations should be consistent with observations. Additionally, a nonzero $\tilde{m}_4^2(t)$, i.e., a contribution from the operator $R^{(3)}\delta g^{00}$, is crucial for avoiding instabilities of scalar perturbations induced by a violation of the NEC \cite{Cai:2016thi,Creminelli:2016zwa,Cai:2017tku}.
However, since we will focus on the primordial GWs in the following calculations, we will not be interested in the explicit formulations of these coefficient functions in the following.

The evolution history of the Universe that we will consider can be divided into three phases, which will be represented by phase 1, 2 and 3. During phase $j$ ($j=1$, 2 or 3), the scale factor could be parameterized with the conformal time $\tau$ as
\be
a_j(\tau)=a_j(\tau_j)\( {\tau -\tau_{R,j} \over \tau_j -\tau_{R,j}} \)^{1\over \epsilon_j-1} \label{a001}
\ee
for $\tau<\tau_j$, where $\tau_j$ is the conformal time at the end of phase $j$, ${\tau}_{R,j} =\tau_{j}-
(\epsilon_{j}-1)^{-1} \mathcal{H}^{-1}(\tau_j)$ is the conformal reference time, $\mathcal{H}(\tau)\equiv a^{-1}d a/d\tau$, $\epsilon_j=-\dot{H}/H^2$ is constant during phase $j$ and $dt=a d\tau$.

The continuities of $a(\tau)$ and $da/d\tau$ are required at $\tau_1$ and $\tau_2$.
Since phase 1 and 3 are assumed as slow-roll inflation while phase 2 is NEC-violating expansion, we will set $\epsilon_1 \approx \epsilon_3\approx 0$ and $\epsilon_2<0$. A specific design of such a model can be found in \cite{Cai:2020qpu}. The detailed variation of $\epsilon$ around the beginning or the end of the NEC-violating phase could be model-dependent. For our purpose, the simplification will not make a qualitative difference.

\subsection{Power spectrum}

With Eqs. (\ref{tensor-action}) and (\ref{QTcT001}), the equation of motion for $\gamma_{\lambda}(\tau,k)$ can be given as
\be \frac{d^{2}
	u_k}{d \tau^{2}}+\left(k^{2}-\frac{d^2a/d\tau^2}{a}\right)
u_k=0\,,\label{eq:eomu} \ee
where we have defined
$u_k=\gamma_{\lambda}(\tau,k)aM_p/2$.
The parameterization in Eq. (\ref{a001}) gives
\be \frac{d^2 a_j/d\tau^2}{a_j}={\nu_j^2-{1/4} \over
	{(\tau-\tau_{R,j})^2}} \label{zppbz02} \ee
with $\nu_j={1\over2}+{1\over 1-\epsilon_j}$.
The solutions to Eq. (\ref{eq:eomu}) during the three phases can be given as
\ba u_{k,1}(\tau)&=&{\sqrt{\pi
		(\tau_{R,1}-\tau)}\over
	2}\lf\{\alpha_{1}H_{\nu_{1}}^{(1)}[k(\tau_{R,1}-\tau)]+\beta_{1}H_{\nu_1}^{(2)}[k(\tau_{R,1}-\tau)]
\rt\} ,\,\, (\tau<\tau_{1})\,,
\\
u_{k,{2}}(\tau)&=&{\sqrt{\pi (\tau_{R,{2}}-\tau)}\over
	2}\Big\{\alpha_{2}H_{\nu_{2}}^{(1)}[k(\tau_{R,2}-\tau)]
+\beta_{2}H_{\nu_{2}}^{(2)}[k(\tau_{R,2}-\tau)] \Big\}
, (\tau_1\leq\tau\leq\tau_{2})\,,
\nn
\\
u_{k,{3}}(\tau)&=&{\sqrt{\pi (\tau_{R,{3}}-\tau)}\over
	2}\Big\{\alpha_{3}H_{\nu_{3}}^{(1)}[k(\tau_{R,3}-\tau)]
+\beta_{3}H_{\nu_{3}}^{(2)}[k(\tau_{R,3}-\tau)] \Big\}
, (\tau_2<\tau \leq \tau_{3})\,,\nn
\label{solution}
\ea
where $\alpha_{j}$ and $\beta_{j}$ are $k$-dependent coefficients for $j=1$, 2, 3;
$H_{\nu_{j}}^{(1)}$ and $H_{\nu_{j}}^{(2)}$ are the first and second kind Hankel functions of the $\nu_j$th order, respectively.

We set the initial state as the Bunch-Davies vacuum (see also
\cite{Piao:2003zm,Piao:2005ag,Liu:2013kea,Qiu:2015nha,Cai:2017pga,Cai:2019hge}), i.e., $u_k= \frac{1}{\sqrt{2 k}
}e^{-i k\tau}$, which indicates $|\alpha_1|=1$ and $|\beta_1|=0$.
Using the matching conditions
$u_{k,j}(\tau_{j})=u_{k,j+1}(\tau_{j})$ and
$d u_{k,j}/d\tau \big|_{\tau=\tau_{j}}=d u_{k,j+1}/d\tau \big|_{\tau=\tau_{j}}$, we can obtain $\alpha_2$, $\beta_2$, $\alpha_3$ and $\beta_3$.
The information of the evolution of the Universe has been encoded in $\alpha_{3}$ and $\beta_{3}$.
For simplicity, we set $\epsilon_1=\epsilon_3=0$, i.e., $\nu_1=\nu_3=3/2$ in the following.
Additionally, we have $1/2<\nu_2<3/2$, since $\epsilon_2<0$.

The resulting power spectrum of primordial GWs is
\be
P_T=\frac{4k^3}{\pi^2
	M_p^2}\cdot\frac{\lf|u_{k,3}
	\rt|^2}{ a^2}= P_{T,inf2} \lf|\alpha_3 - \beta_3
\rt|^2\,,\label{eq:PT}
\ee
where $P_{T,inf2}={2H_{inf2}^2 \over M_p^2 \pi^2}$, $H_{inf2}=H(\tau_3)$ is the Hubble parameter at the second inflation phase (i.e., phase 3).
With the matching method, we obtain that
\ba
\alpha_3-\beta_3 &=&
{\pi(1-2\nu_2)\over 8}\sqrt{x_1 y_1} \left\{H^{(1)}_{\nu_2}(y_2)\left[(i-{1\over x_1})H^{(2)}_{\nu_2-1}(x_2) + H^{(2)}_{\nu_2}(x_2)  \right]\sin y_1  \right.
\nn\\
&\,&
+ H^{(1)}_{\nu_2-1}(y_2) \left[(i-{1\over x_1})H^{(2)}_{\nu_2-1}(x_2) + H^{(2)}_{\nu_2}(x_2)   \right] \left[\cos y_1 - {1\over y_1}\sin y_1 \right]
\nn\\
&\,&
\left. -\left[(i-{1\over x_1})H^{(1)}_{\nu_2-1}(x_2) + H^{(1)}_{\nu_2}(x_2) \right] \left[ H^{(2)}_{\nu_2}(y_2)\sin y_1 + H^{(2)}_{\nu_2-1}(y_2)(\cos y_1 - {1\over y_1}\sin y_1) \right] \right\} \,,
\label{eq:alphabeta01}
\nn\\
\ea
where $x_1=k/\bar{\cal H}_1$, $x_2=(\nu_2-1/2)x_1$, $y_1=k/\bar{\cal H}_2$, $y_2=(\nu_2-1/2)y_1$, $\bar{\cal H}_1={\cal H}(\tau_1)$ and $\bar{\cal H}_2={\cal H}(\tau_2)$; a phase factor has been neglected.
It should be emphasized again that $a$ and $d a/d\tau$ are required to be continuous at $\tau_1$ and $\tau_2$, which is reasonable. In order to obtain the asymptotic behavior of $P_T/P_{T,inf2}$ with respect to the comoving wavenumber $k$, let us assume that the NEC-violating phase lasts a sufficient long time such that $\bar{\cal H}_1\ll \bar{\cal H}_2$.

For perturbations that exit the horizon in the first inflationary phase, we have $k\ll \bar{\cal H}_1\ll\bar{\cal H}_2$, i.e., $x_1$, $x_2$, $y_1$, $y_2\ll1$, provided $1/2<\nu_2<3/2$. By straightforward series expansion, we find
\ba
{P_T\over P_{T,inf2}}\Big|_{k\ll \bar{\cal H}_1} = \lf({\bar{\cal H}_1\over \bar{\cal H}_2} \rt)^{3-2\nu_2}+{\cal O}\( {k\over \bar{\cal H}_1}\)^2 \,. \label{eq:PTNEC01}
\ea
As can be inferred, the power spectrum is (nearly) scale-invariant in the limit of $k\ll \bar{\cal H}_1$.
Note that, according to Eq. (\ref{a001}), the leading order term in the right hand side of Eq. (\ref{eq:PTNEC01}) can also be written as $({\bar{\cal H}_1/ \bar{\cal H}_2} )^{3-2\nu_2} = ({H_1/ H_2} )^{2}$, where $H_1$ and $H_2$ are evaluated at $\tau_1$ and $\tau_2$, respectively.

A similar calculation gives ${P_T\over P_{T,inf2}}\Big|_{k\ll \bar{\cal H}_1} \simeq {4\over 9}\lf({\bar{\cal H}_1\over \bar{\cal H}_2} \rt)^{3-2\nu_2}$ for $\nu_2=1/2$, where there is a difference of a factor $4/9$ at the leading order compared with (\ref{eq:PTNEC01}). However, it should be pointed out that when $\nu_2=1/2$, i.e., $\epsilon_2\rightarrow -\infty$, the parameterization given by Eq. (\ref{a001}) cannot guarantee the continuity of $d a/d\tau$ at $\tau_1$ or $\tau_2$ any more. The above derivation will no longer be applicable. Therefore, we will disregard the case of $\nu_2=1/2$ in this paper.

For perturbations that exit the horizon in the NEC-violating phase, we will focus on the perturbation modes which satisfy $\bar{\cal H}_1 \ll k\ll \bar{\cal H}_2$, i.e., $x_{1}$, $x_2 \gg1$ and $y_{1}$, $y_2\ll1$. We obtain
\ba
{P_T\over P_{T,inf2}}\Big|_{\bar{\cal H}_1 \ll k\ll \bar{\cal H}_2} = {\Gamma^2(\nu_2)\over \pi} \({2\nu_2-1 \over 4}  \)^{1-2\nu_2}\( {k\over \bar{\cal H}_2}\)^{3-2\nu_2}
+ {\cal O}\[{\bar{\cal H}_1^2 \over \bar{\cal H}_2^2} \( {k \over \bar{\cal H}_2}\)^{1-2\nu_2} \] \label{eq:PTNEC02}
\ea
for $1/2<\nu_2<3/2$.
Therefore, the GWs generated during the NEC-violating phase have a blue spectrum, where the spectrum index is $n_T=3-2\nu_2$.

As for the perturbation modes that exit the horizon in phase 3, namely, the second inflationary phase, we have $k\gg \bar{\cal H}_2$, which indicates $x_{1}$, $x_2 \gg1$ and $y_{1}$, $y_2\gg 1$. The result is trivially given as
\ba
{P_T\over P_{T,inf2}}\Big|_{k\gg \bar{\cal H}_2} = 1 +{\cal O}\( {k\over \bar{\cal H}_2}\)^{-2}  \,, \label{eq:PTNEC03}
\ea
where there is some oscillating corrections included in the subleading order terms.

Apparently, when $\nu_2\rightarrow 3/2$, all of Eqs. (\ref{eq:PTNEC01}) to (\ref{eq:PTNEC03}) reduce to ${P_T/ P_{T,inf2}}\rightarrow 1$, which is just the result of a single inflationary phase.
In a realistic inflation model which is constructed with a slowly rolling scalar field, the slow roll parameter $\epsilon$ should slightly deviate from $0$. The resulting power spectrum would also deviate from scale-invariance. We assume that such a deviation could be absorbed into $P_{T,inf2}$, for simplicity.

\subsection{Templates}

\begin{figure}[htbp]
    \subfigure[~~]{\includegraphics[width=.47\textwidth]{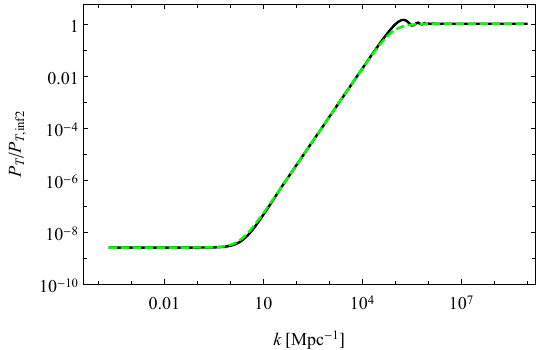} }
    \subfigure[~~]{\includegraphics[width=.47\textwidth]{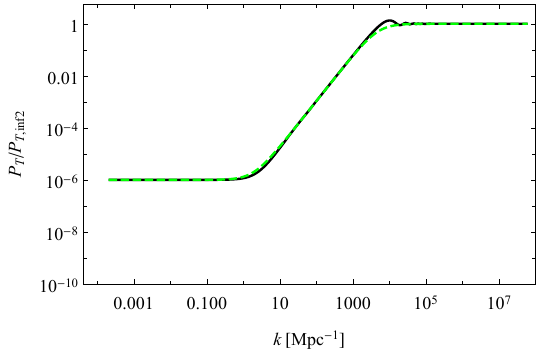} }
    \caption{Comparisons between the template (\ref{eq:para01}) (green dashed curve) and the numerical result of ${P_T/ P_{T,inf2}}$ (black solid curve). We have set $\bar{\cal H}_1=10^4a_0H_0$, where $a_0=1$ and $H_0=67\, {\rm km/s/Mpc}$. Left: $\epsilon_2=-13.5$, $\bar{\cal H}_2=4.165\times 10^4 \bar{\cal H}_1$. Right: $\epsilon_2=-7.5$, $\bar{\cal H}_2=2.512\times 10^3 \bar{\cal H}_1$. } \label{fig01}
\end{figure}

With the asymptotic behaviors of ${P_T/ P_{T,inf2}}$ given by Eqs. (\ref{eq:PTNEC01}) to (\ref{eq:PTNEC03}), we can provide a template of the GW power spectrum in our scenario as
\ba
{P_T(k)\over P_{T,inf2}} = {
\lf(\bar{\cal H}_1\over \bar{\cal H}_2 \rt)^{3-2\nu_2}
+ {\Gamma^2(\nu_2)\over \pi} \({2\nu_2-1 \over 4}  \)^{1-2\nu_2}\( {k\over \bar{\cal H}_2}\)^{3-2\nu_2}
\over
1
+ {\Gamma^2(\nu_2)\over \pi} \({2\nu_2-1 \over 4}  \)^{1-2\nu_2}\( {k\over \bar{\cal H}_2}\)^{3-2\nu_2}
}\,, \label{eq:para01}
\ea
where there are only three model-dependent parameters, i.e., $\nu_2 \equiv {1\over2}+{1\over 1-\epsilon_2}$, $\bar{\cal H}_1$ and $\bar{\cal H}_2$. Here, $\bar{\cal H}_1$ and $\bar{\cal H}_2$ can also be replaced by $H_1$ and $H_2$, since $({\bar{\cal H}_1/ \bar{\cal H}_2} )^{3-2\nu_2} = ({H_1/ H_2} )^{2}$. The condition $\bar{\cal H}_1\ll \bar{\cal H}_2$ or $H_1\ll H_2$ is required for the template to be applicable.
Obviously, in the limit $k\ll \bar{\cal H}_1$, $\bar{\cal H}_1 \ll k\ll \bar{\cal H}_2$ and $k\gg \bar{\cal H}_2$, Eq. (\ref{eq:para01}) reduces to Eq. (\ref{eq:PTNEC01}), (\ref{eq:PTNEC02}) and (\ref{eq:PTNEC03}) up to the leading order, respectively. A comparison between the template (\ref{eq:para01}) and the numerical result of ${P_T/ P_{T,inf2}}$ is displayed in Fig. \ref{fig01}.

\begin{figure}[htbp]
    \subfigure[~~]{\includegraphics[width=.47\textwidth]{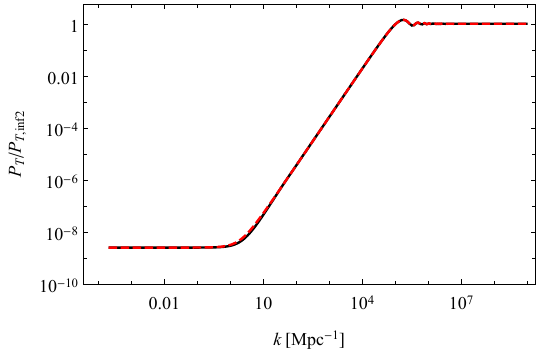} }
    \subfigure[~~]{\includegraphics[width=.47\textwidth]{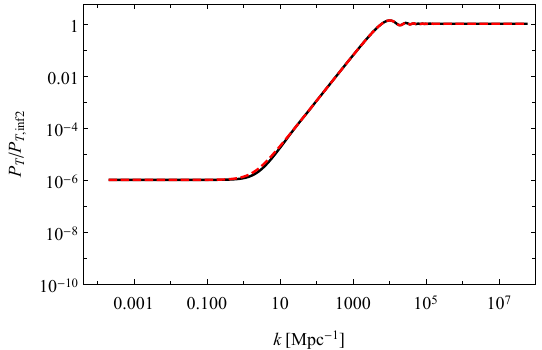} }
    \caption{Comparisons between the template (\ref{eq:para02}) (green dashed curve) and the numerical result of ${P_T/ P_{T,inf2}}$ (black solid curve). We have set $\bar{\cal H}_1=10^4a_0H_0$, where $a_0=1$ and $H_0=67\, {\rm km/s/Mpc}$. Left: $\epsilon_2=-13.5$, $\bar{\cal H}_2=4.165\times 10^4 \bar{\cal H}_1$. Right: $\epsilon_2=-7.5$, $\bar{\cal H}_2=2.512\times 10^3 \bar{\cal H}_1$. } \label{fig02}
\end{figure}

The parameterization of the template (\ref{eq:para01}) is quite simple and straightforward.
However, a shortcoming of (\ref{eq:para01}) is the loss of the oscillating features in ${P_T/ P_{T,inf2}}$ around $k\sim \bar{\cal H}_1$ and $k\sim \bar{\cal H}_2$. These oscillations could be model-dependent to some extent. Additionally, since $\bar{\cal H}_1\ll \bar{\cal H}_2$ in our model, the oscillations around $k\sim \bar{\cal H}_1$ are suppressed greatly compared with those around $k\sim \bar{\cal H}_2$, as can be seen from Eq. (\ref{eq:alphabeta01}), see also Fig. \ref{fig01}. Therefore, only the oscillations around $k\sim \bar{\cal H}_2$ need to be included in the template.
For this purpose, we can use the condition $k\gg \bar{\cal H}_1$ without considering the relation between $k$ and $\bar{\cal H}_2$. The template can be given as
\ba
{P_T(k)\over P_{T,inf2}} = \lf(\bar{\cal H}_1\over \bar{\cal H}_2 \rt)^{3-2\nu_2}
&+&
{\pi (2 \nu_ 2 - 1) \over 4} {k\over \bar {\cal H} _ 2} {\rm Abs}\left\{ H^{(1)}_{\nu_2}\[(\nu_2-{1\over2}){k\over \bar{\cal H}_2}\]\sin {k\over \bar{\cal H}_2} \right.
\nn\\ \qquad\qquad\quad\,
&+& \left. H^{(1)}_{\nu_2-1}\[(\nu_2-{1\over2}){k\over \bar{\cal H}_2}\]\[\cos{k\over \bar{\cal H}_2}-{\bar{\cal H}_2 \over k}\sin {k\over \bar{\cal H}_2} \] \right\}^2\,, \label{eq:para02}
\ea
where there are still three model-dependent parameters, i.e., $\nu_2$, $\bar{\cal H}_1$ and $\bar{\cal H}_2$.
It is obvious that Eq. (\ref{eq:para02}) reduces to (\ref{eq:PTNEC01}) in the limit $k\ll \bar{\cal H}_1$. For $k \gtrsim \bar{\cal H}_1$, Eq. (\ref{eq:para02}) is equivalent to (\ref{eq:alphabeta01}) at the leading order, so that the oscillating features around $k\ll \bar{\cal H}_2$ is preserved, see Fig. \ref{fig02}.

\section{Enhanced power spectrum of primordial GWs from the diminishment of propagating speed of inflationary GWs}
\label{sec:cTdim}

With a disformal transformation of the metric, a time-dependent $c_T$ can be reset as unity \cite{Creminelli:2014wna}. However, the background evolution may no longer be inflationary. For example, it has been found that slow-roll inflation in which $c_T$ decreases rapidly with time is a disformal dual to a superinflation ($\epsilon<0$) in which $c_T\equiv 1$ \cite{Cai:2015yza}.
The power spectrum, which is observable in principle, has to be invariant under a nonsingular disformal transformation.
Therefore, a time-dependent $c_T$ during inflation is able to imprint interesting features in the power spectra of primordial GWs \cite{Cai:2015dta,Cai:2015ipa,Cai:2015yza,Cai:2016ldn}.\footnote{See also \cite{Giovannini:2018zbf,Giovannini:2018nkt,Giovannini:2021uvh} for the refractive index of the relic gravitons.}
In this section, we calculate the enhanced power spectra of primordial GWs for a specific time-dependent $c_T$ while assuming the background evolution as the slow-roll inflation.
%We also discuss the relation between such a scenario and the one studied in Sec. \ref{Sec:NECV01}.

\subsection{Set up}\label{sec:cTsetup}

Using the EFT method, we can write the action as
\ba
S&=&\int
d\hat{t}d^3\hat{x}\sqrt{-\hat{g}}\Big[ {M_p^2\over2} \hat{R}-\hat{\Lambda}(t)-\hat{c}(t)\hat{g}^{00}
\nn\\
&\,&\qquad\qquad\quad
-\hat{m}_4^2(t)\lf( \delta \hat{K}^2-\delta \hat{K}_{\mu\nu}\delta
\hat{K}^{\mu\nu} \rt)
\Big] \,,
\label{action03}
\ea
where ` $\hat{}$ 's are used to mark the frame in which the propagating speed of GWs is time-dependent.
We assume that the background evolution is the slow-roll inflation throughout, which is determined by the first line of action (\ref{action03}). Again, we have neglected $S_{\rm m}$. However, it should be remembered that the matter sector is crucial for identifying the physical background evolution as measured relative to physical rulers, see e.g., \cite{Ijjas:2015zma}.
Additionally, operators $(\delta \hat{g}^{00})^2$, $\delta \hat{K}\delta \hat{g}^{00}$ and $\hat{R}^{(3)}\delta \hat{g}^{00}$ can also appear in the action (\ref{action03}), since they will not affect the tensor perturbations at quadratic order, as can be seen from Eq. (\ref{eq:QTcT01}). For simplicity, we will focus on the tensor perturbations and consider only the contribution of a nonzero $\hat{m}_4^2(t)$, which contributes nothing to the scalar perturbation at quadratic order.

According to Eqs. (\ref{action01}) to (\ref{eq:QTcT01}), we can write the quadratic action of $\gamma_{ij}$ as
\be
S^{(2)}_{\gamma}={M_p^2\over8}\int d\hat{\tau}d^3x \hat{a}^2
\hat{Q}_T\lf[ \({d \gamma_{ij}\over d\hat{\tau}} \)^2 -\hat{c}_T^2(\partial_k\gamma_{ij})^2 \rt]\,, \label{tensor-action-cT}
\ee
where
\be
\hat{Q}_T=\hat{c}_T^{-2},\qquad \hat{c}_T^2=\(1+{2 \hat{m}_4^2 \over M_p^2}\)^{-1}\,. \label{QTcT002}
\ee

As in the last section, we consider three phases, during which the scale factor can be parameterised as
\be
\hat{a}_j(\hat{\tau})= \hat{a}_j(\hat{\tau}_j)\( {\hat{\tau} -\hat{\tau}_{R,j} \over \hat{\tau}_j -\hat{\tau}_{R,j}} \)^{1\over \hat{\epsilon}_j-1} \label{a002}
\ee
for $\hat{\tau}<\hat{\tau}_j$, where $\hat{\tau}_j$ is the conformal time at the end of phase $j$, $\hat{\tau}_{R,j} =\hat{\tau}_{j}-
(\hat{\epsilon}_{j}-1)^{-1} \hat{\mathcal{H}}^{-1}(\hat{\tau}_j)$, $\hat{\mathcal{H}}(\hat{\tau})\equiv \hat{a}^{-1}d \hat{a}/d\hat{\tau}$. Since the background evolution is assumed as slow-roll inflation throughout these three phases, we will set $\hat{\epsilon}_1=\hat{\epsilon}_2=\hat{\epsilon}_3=0$ in the following, for simplicity.

The propagating speed of primordial GWs can be parameterized as
\ba
\hat{c}_{T,1}(\hat{\tau})&=&\bar{\hat{c}}_{T,1}:=\hat{c}_{T,2}(\hat{\tau}_1)\,,\qquad\qquad  (\hat{\tau} < \hat{\tau}_{1})\,, \label{eq:cT01}
\\
\hat{c}_{T,2}(\hat{\tau})&=&\bar{\hat{c}}_{T,2}\( {\hat{\tau} -\hat{\tau}_{R,2} \over \hat{\tau}_2 -\hat{\tau}_{R,2}} \)^p \,, \qquad ~\, (\hat{\tau}_{1} \leq \hat{\tau} \leq \hat{\tau}_{2})\,,\label{eq:cT02}
\\
\hat{c}_{T,3}(\hat{\tau})&=&\bar{\hat{c}}_{T,2}:=\hat{c}_{T,2}(\hat{\tau}_2) \,, \qquad \qquad  (\hat{\tau}_{2} < \hat{\tau} \leq \hat{\tau}_{3})\,, \label{eq:cT03}
\ea
where the parameter $p={\rm const.}>0$, $\bar{\hat{c}}_{T,1}>\bar{\hat{c}}_{T,2}$. Namely, we assume that $\hat{c}_{T}(\hat{\tau})$ is constant during phase 1 and phase 3, while $\hat{c}_{T}(\hat{\tau})$ decreases with time during phase 2.

\subsection{Power spectrum}

In order to analytically calculate the power spectrum induced by a time-dependent $\hat{c}_{T}(\hat{\tau})$, we define
\be
d\sigma=\hat{c}_{T}(\hat{\tau}) d\hat{\tau}\,, \label{eq:ytau01}
\ee
so that the quadratic action (\ref{tensor-action-cT}) can be rewritten as
\be
S^{(2)}_{\gamma}={M_p^2\over8}\int d\sigma d^3x \hat{a}^2
\hat{c}_T^{-1}\lf[ \({d \gamma_{ij}\over d\sigma} \)^2 -(\partial_k\gamma_{ij})^2 \rt]\,. \label{tensor-action-cT-2}
\ee
The equation of motion for $\gamma_{\lambda}(\tau,k)$ is
\be \frac{d^{2}
	v_k}{d \sigma^{2}}+\left(k^{2}-\frac{d^2 \hat{z}_T/d \sigma^2}{\hat{z}_T}\right)
v_k=0\,,\label{eq:eomu-cT-02}
\ee
where $v_k=\gamma_{\lambda}(\tau,k)\hat{z}_T$, $\hat{z}_T=M_p \hat{a}\hat{c}_T^{-1/2}/2$.
Using Eq. (\ref{eq:PT01}), the power spectrum can be given as
\be
\hat{P}_T={4k^3 \over \pi^2 M_p^2 }{\hat{c}_T \over \hat{a}^2}|v_k|^2\,. \label{eq:PTcT01}
\ee

From Eq. (\ref{eq:cT01}) to (\ref{eq:cT03}), we can write the solutions to Eq. (\ref{eq:eomu-cT-02}) in the three phases as
\ba \label{eq:vk01}
v_{k,1}(\sigma)&=& {\sqrt{\pi(\sigma_{R,1}-\sigma)}\over 2 }\lf\{\hat{\alpha}_1 H^{(1)}_{\hat{\nu}_1}[k(\sigma_{R,1}-\sigma)] + \hat{\beta}_1H^{(2)}_{\hat{\nu}_1}[k(\sigma_{R,1}-\sigma)]  \rt\}, (\sigma<\sigma_1)\,,
\\
v_{k,2}(\sigma)&=& {\sqrt{\pi(\sigma_{R,2}-\sigma)}\over 2 }\lf\{\hat{\alpha}_2 H^{(1)}_{\hat{\nu}_2}[k(\sigma_{R,2}-\sigma)] + \hat{\beta}_2 H^{(2)}_{\hat{\nu}_2}[k(\sigma_{R,2}-\sigma)]  \rt\}, (\sigma_1\leq \sigma \leq \sigma_2)\,,
\nn\\
v_{k,3}(\sigma)&=& {\sqrt{\pi(\sigma_{R,3}-\sigma)}\over 2 }\lf\{\hat{\alpha}_3 H^{(1)}_{\hat{\nu}_3}[k(\sigma_{R,3}-\sigma)] + \hat{\beta}_3 H^{(2)}_{\hat{\nu}_3}[k(\sigma_{R,3}-\sigma)]  \rt\}, (\sigma_2< \sigma \leq \sigma_3)\,,\nn
\ea
where
\be
\hat{\nu}_1=\hat{\nu}_3=3/2\,,\qquad \hat{\nu}_2={3+2p\over 2+2p}\,,
\ee
$\sigma_{R,j}$ corresponds to $\hat{\tau}_{R,j}$ according to the relation (\ref{eq:ytau01}), $\hat{\alpha}_j$ and $\hat{\beta}_j$ are $k$-dependent coefficients.
The initial condition of $v_k$ is set as $v_k\simeq {1\over \sqrt{2k}}e^{-ik \sigma}$, which indicates that $|\hat{\alpha}_1|=1$ and $|\hat{\beta}_1|=0$.

With Eqs. (\ref{eq:PTcT01}) and (\ref{eq:vk01}), the power spectrum evaluated at the end of phase 3 can be given as
\be
\hat{P}_T=\hat{P}_{T,inf2}|\hat{\alpha}_3-\hat{\beta}_3|^2\,, \label{eq:PTcTab-1}
\ee
where $\hat{P}_{T,inf2}={2 \hat{H}^2_{3} \over \pi^2 M_p^2 \bar{\hat{c}}_{T,2} }$, $\hat{H}_{j}$ is the Hubble parameter evaluated at $\hat{\tau}_j$. Since we have assumed that $\hat{\epsilon}_3=0$, we have $\hat{H}_{3}=\hat{H}_{2}=\hat{H}_{inf2}$.
For a specific model of slow-roll inflation construct by scalar-tensor theories, the slow-roll parameter $\hat{\epsilon}_j$ can slightly deviate from 0. Therefore, there could be a correction in $\hat{P}_{T,inf2}$, which will be ignored in this paper for simplicity.

We can obtain $\hat{\alpha}_3$ and $\hat{\beta}_3$ by requiring the continuities of $v_k$ and $d v_k/d\sigma$ at $\sigma_1$ and $\sigma_2$.
Note that the continuities of $\hat{\cal H}$ at $\hat{\tau}_1$ and $\hat{\tau}_2$ actually requires that $\hat{\tau}_{R,1}=\hat{\tau}_{R,2}=\hat{\tau}_{R,3}$. Hence, we also have $\sigma_{R,1}=\sigma_{R,2}=\sigma_{R,3}=:\sigma_{R}$.
For convenience, we define $z_1=k(\sigma_{R}-\sigma_{1})$ and $z_2=k(\sigma_{R}-\sigma_{2})$.
Using (\ref{eq:ytau01}), it can be found that $z_1\approx {k\over (1+p)\hat{\cal H}_1/\bar{\hat{c}}_{T,1}}$ and $z_2={k\over \hat{\cal H}_2/\bar{\hat{c}}_{T,2}}$.
%Instead of writing down the full formulae of $\hat{\alpha}_3$ and $\hat{\beta}_3$, it is more useful to provide the asymptotic behaviors of $|\hat{\alpha}_3-\hat{\beta}_3|^2$.
As a result, we obtain
\ba
\hat{\alpha}_3-\hat{\beta}_3 &=&
{-\pi\over 16 z_ 1^{3/2} z_ 2^{3/2}} \left\{
2 z_2 H^{(1)}_{\hat{\nu}_2-1}(z_2) \left[2 i z_1 (i + z_1) H^{(2)}_{\hat{\nu}_2 -1} (z_1) \right. \right.
\nn\\
&\,& \left. + (2 \hat{\nu}_2 - 3 +z_1 (3 i + 2 z_1 -2 i \hat{\nu}_2)) H^{(2)}_{\hat{\nu}_2}(z_1) \] \
(z_2 \cos z_2 - \sin z_2 )
\nn\\
&\,&\, + \lf[ z_2 (3 - 2 \hat{\nu}_2) \cos z_2  + ( 2 \hat{\nu}_2-3 + 2 z_2^2  ) \sin z_2 \rt] H^{(1)}_{\hat{\nu}_2}(z_2)\lf[ 2i z_1 (i+z_1)H^{(2)}_{\hat{\nu}_2 -1} (z_1) \right.
\nn\\
&\,&
\left. +  (2 \hat{\nu}_2 - 3 +z_1 (3 i + 2 z_1 -2 i \hat{\nu}_2)) H^{(2)}_{\hat{\nu}_2}(z_1)  \rt]
\nn\\
&\,&
\, + \lf[2i z_1(i+z_1)H^{(1)}_{\hat{\nu}_2 -1} (z_1) + (2 \hat{\nu}_2 - 3 + z_1 (3 i + 2 z_1 -2 i \hat{\nu}_2)) H^{(1)}_{\hat{\nu}_2}(z_1) \rt]
\nn\\
&\,&
\,\,\, \cdot \left[ 2z_2 H^{(2)}_{\hat{\nu}_2 -1}(z_2)(\sin z_2-z_2\cos z_2)  \right.
\nn\\
&\,&
\left.  \left. + H^{(2)}_{\hat{\nu}_2}(z_2)\lf[z_2(2\hat{\nu}_2-3)\cos z_2 + 3\sin z_2 -2(z_2^2+\hat{\nu}_2)\sin z_2  \rt]
\right]
\right\}\,.
\ea

For the perturbation modes exited horizon during phase 1 (i.e., $\sigma<\sigma_1$), we have $z_1\ll1$ and $z_2\ll1$. Therefore, we find that
\ba
{\hat{P}_T\over \hat{P}_{T,inf2}}\Big|_{k\ll \hat{\cal H}_1/\bar{\hat{c}}_{T,1}} \simeq \frac{\bar{\hat{c}}_{T,2}}{\bar{\hat{c}}_{T,1}} + {\cal O}\({k\over \hat{\cal H}_1/\bar{\hat{c}}_{T,1}}\)^2 \,. \label{eq:PTcT02}
\ea
%\ba
%{P_T\over P_{T,3}}\Big|_{k\ll \hat{\cal H}_1/\bar{\hat{c}}_{T,1}} =\frac{\bar{\hat{c}}_{T,2}}{\bar{\hat{c}}_{T,1}} \frac{(p+1)^{-\frac{p+2}{p+1}} }{144 (2 p+3)^2} \left[(5 p+6)^2-p^2 (p+1)^{\frac{1}{p+1}+2} \left(\frac{\bar{\hat{c}}_{T,2}}{\bar{\hat{c}}_{T,1}}\right)^{\frac{3}{p}+2}\right]^2  \,. \label{eq:PTcT01}
%\ea
where we have used $\hat{\cal H}_2/\hat{\cal H}_1=(\bar{\hat{c}}_{T,1}/\bar{\hat{c}}_{T,2})^{1/p}$, which can be obtained from the continuity of $\hat{c}_T$ at $\hat{\tau}_1$.
Note that there is actually a $p$-dependent factor in the leading order term of the right hand side of Eq. (\ref{eq:PTcT02}), which appears due to the fact that $d\hat{c}_{T}(\hat{\tau})/d\hat{\tau}$ is not continuous at $\hat{\tau}_1$ or $\hat{\tau}_2$ according to Eqs. (\ref{eq:cT01}) to (\ref{eq:cT03}). Such issues also appear in Sec. \ref{Sec:NECV01} for the case where $\nu_2=1/2$.
In fact, for ${k\ll \hat{\cal H}_1/\bar{\hat{c}}_{T,1}}$, we can directly obtain the power spectrum by substituting $v_{k,1}$ into Eq. (\ref{eq:PTcT01}), which gives $\hat{P}_{T,inf1}={2\hat{H}_1^2 \over \pi^2 M_p^2 \bar{\hat{c}}_{T,1}}$, i.e., $\hat{P}_{T,inf1}/ \hat{P}_{T,inf2}= {\bar{\hat{c}}_{T,2}/\bar{\hat{c}}_{T,1}}$, provided these perturbation modes are frozen after horizon crossing.

For the perturbation modes exited horizon during phase 2 (i.e., $\sigma_1\leq \sigma \leq \sigma_2$), we consider the situation in which $z_1\gg1$ and $z_2\ll1$. Then, we find
\ba
{\hat{P}_T\over \hat{P}_{T,inf2}}\Big|_{\hat{\cal H}_1/\bar{\hat{c}}_{T,1} \ll k\ll \hat{\cal H}_2/\bar{\hat{c}}_{T,2} }
\simeq {2^{2 \hat{\nu}_2-3} (3 + 2 \hat{\nu}_2)^2 \Gamma^2 (\hat{\nu}_2) \over 9\pi}\lf( {k\over \hat{\cal H}_2/\bar{\hat{c}}_{T,2}} \rt)^{3 - 2 \hat{\nu}_2}
+ {\cal O}\lf( {k\over \hat{\cal H}_1/\bar{\hat{c}}_{T,1}} \rt)^{ 2 \hat{\nu}_2-5}
\,. \nn\\
\label{eq:PTcT03}
\ea
It can be checked that $\hat{P}_T({\hat{\cal H}_1/\bar{\hat{c}}_{T,1} \ll k\ll \hat{\cal H}_2/\bar{\hat{c}}_{T,2} })\rightarrow \hat{P}_{T,inf2}$ in the limit $p\rightarrow 0$.
Again, the discontinuity of $d\hat{c}_{T}(\hat{\tau})/d\hat{\tau}$ at $\hat{\tau}_1$ or $\hat{\tau}_2$ could contribute a $p$-dependent factor, which is not physical. What we shall be interested in is the spectrum index, i.e., $\hat{n}_{T,2}=3 - 2 \hat{\nu}_2 = p/(1+p)$, which is consistent with that obtained in \cite{Cai:2015yza,Cai:2016ldn}.

\begin{figure}
	\centering
	\includegraphics[angle=0, width=0.6\textwidth]{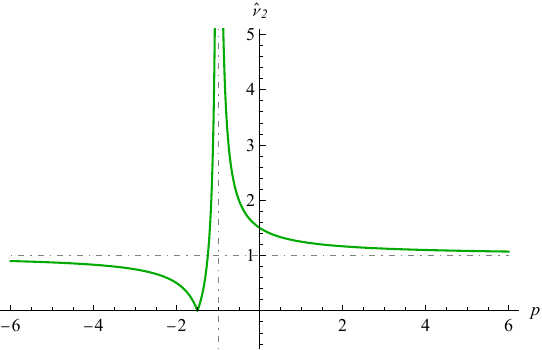}
	\caption{For a general $p$, we have $\hat{\nu}_2=\lf|{3+2p\over 2+2p}\rt|$, which is depicted by the green curve.}
	\label{fig:nu2-p}
\end{figure}

Note that we have assumed $p>0$ in the derivation of Eq. (\ref{eq:PTcT03}). For a general $p$, we actually have $\hat{\nu}_2=\lf|{3+2p\over 2+2p}\rt|$, which is displayed in Fig. \ref{fig:nu2-p}. It can be checked that $\hat{n}_{T,2}=3 - 2 \hat{\nu}_2 = p/(1+p)$ is also valid for an increasing $c_T$ as long as $-1<p\leq 0$, which will predict a red-tilted (i.e., $\hat{n}_{T,2}<0$) spectrum of primordial GWs. It is interesting to see whether a blue spectrum can be obtained by an increasing $c_T$ with $p<-1$.

\begin{figure}
	\centering
	\includegraphics[angle=0, width=0.95\textwidth]{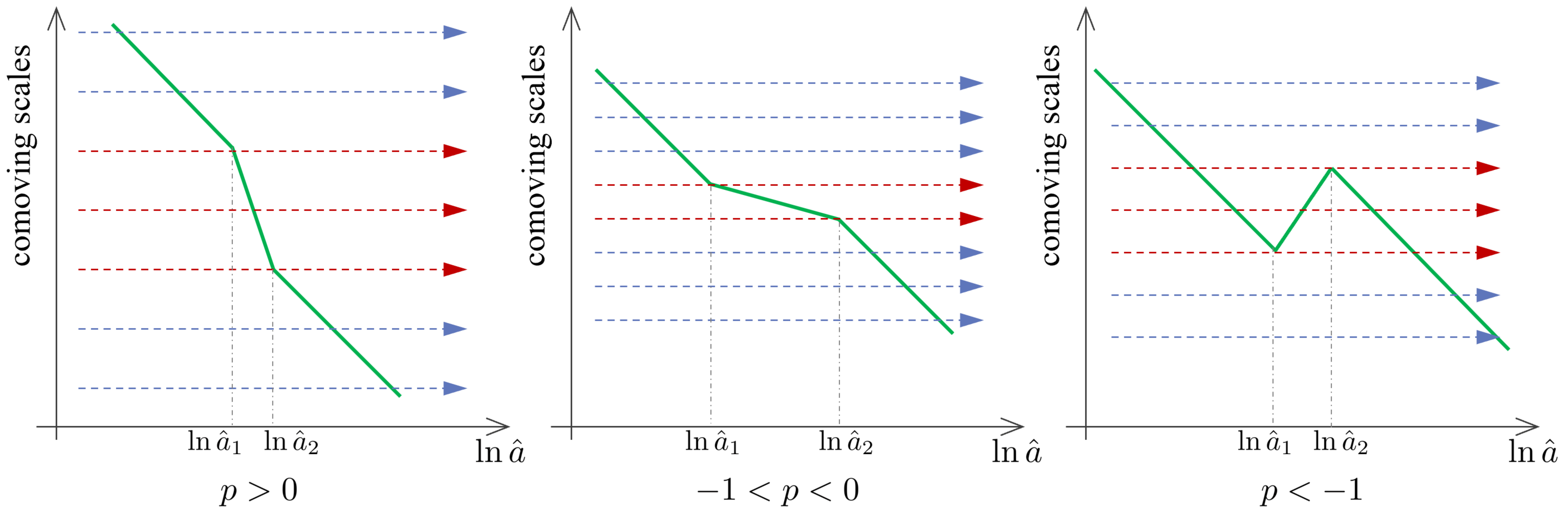}
	\caption{Horizon-crossing of primordial GWs with time-dependent propagating speed $c_T$ parameterized by Eqs. (\ref{eq:cT01}) to (\ref{eq:cT03}). The green lines represent the evolution of the effective comoving sound horizon of GWs, which is defined as ${1\over a|H_{\rm GW}|}\simeq \sqrt{\frac{\hat{z}_T}{d^2 \hat{z}_T/d \sigma^2}}\simeq \hat{c}_T/\hat{\cal H}$; the dashed lines with arrows represent the GW modes, in which the modes with comoving wavenumber $\hat{\cal H}_1/\bar{\hat{c}}_{T,1} < k < \hat{\cal H}_2/\bar{\hat{c}}_{T,2}$ for $p>-1$ (and $\hat{\cal H}_1/\bar{\hat{c}}_{T,1} > k> \hat{\cal H}_2/\bar{\hat{c}}_{T,2}$ for $p<-1$) are colored red. During the stage where $c_T$ increases with time while $p<-1$, the GW modes cannot exit their sound horizon.}
	\label{fig:Horizon-p}
\end{figure}

Unfortunately, the result of Eq. (\ref{eq:PTcT03}) is no longer applicable for the situation $p<-1$.
Actually, according to Eqs. (\ref{a002}), (\ref{eq:cT02}) and (\ref{eq:eomu-cT-02}), the effective comoving sound horizon of tensor perturbation mode, which can be defined as ${1\over a|H_{\rm GW}|}\equiv\sqrt{\lf|\frac{\hat{z}_T}{d^2 \hat{z}_T/d \sigma^2} \rt|}\sim {\hat{c}_T\over\hat{\cal H}}\sim |\hat{\tau} -\hat{\tau}_{R,2}|^{p+1}$, will not decrease with time when $p\leq -1$. In other words, we will have ${\hat{\cal H}_1/\bar{\hat{c}}_{T,1} \gg \hat{\cal H}_2/\bar{\hat{c}}_{T,2} }$ in Eq. (\ref{eq:PTcT03}) for $p<-1$.
Therefore, the tensor perturbation modes are not able to exit their sound horizon when $p\leq -1$. In fact, the tensor perturbation modes with comoving wavenumber ${\hat{\cal H}_1/\bar{\hat{c}}_{T,1} \gg k\gg \hat{\cal H}_2/\bar{\hat{c}}_{T,2} }$ should reenter their sound horizon during the $p<-1$ phase and eventually exit their horizon during the third phase ($\hat{\tau}_{2} < \hat{\tau} \leq \hat{\tau}_{3}$), see Fig. \ref{fig:Horizon-p} for an illustration. With this in mind, we still obtain a red-tilted spectrum, i.e., ${\hat{P}_T / \hat{P}_{T,inf2}} \sim k^{-(3 + 2 \hat{\nu}_2)}$, for ${\hat{\cal H}_1/\bar{\hat{c}}_{T,1} \gg k\gg \hat{\cal H}_2/\bar{\hat{c}}_{T,2} }$. In a short summary, the case in which $c_T$ increases with time (i.e., $p<0$) during inflation cannot generate GWs with a blue spectrum.

For the perturbation modes exited horizon during phase 3 (i.e., $\sigma_2< \sigma \leq \sigma_3$), we have $z_1\gg1$ and $z_2\gg1$. In this limit, the effect of the discontinuity of $d\hat{c}_{T}(\hat{\tau})/d\hat{\tau}$ can be negligible, as can be inferred from Eq. (\ref{eq:eomu-cT-02}). Therefore, we can obtain that
\ba
{\hat{P}_T\over \hat{P}_{T,inf2}}\Big|_{k\gg \hat{\cal H}_2/\bar{\hat{c}}_{T,2}} = 1+ {\cal O}\lf( {k\over \hat{\cal H}_2/\bar{\hat{c}}_{T,2}} \rt)^{ -2 } \,, \label{eq:PTcT04}
\ea
where the leading order term is trivial.

When $\hat{\tau} > \hat{\tau}_{3}$, $\hat{c}_{T}$ may regain the value of the speed of light, so that the predictions will be consistent with observations of GWs at higher frequency \cite{LIGOScientific:2018dkp}. As a consequence, the power spectrum of GWs could become red-tilted at much higher frequencies.
Additionally, the discontinuity of $d\hat{c}_{T}(\hat{\tau})/d\hat{\tau}$ at $\hat{\tau}_1$ or $\hat{\tau}_2$ is a defect of the parameterizations (\ref{eq:cT01}) to (\ref{eq:cT03}).
For a generic smooth $\hat{c}_{T}(\hat{\tau})$, the power spectrum can be solved numerically from the action (\ref{tensor-action-cT}), see the next section for two examples.

\section{Discussion of relation and discrepancy between two scenarios}\label{Sec:relation01}

Both actions (\ref{action02}) and (\ref{action03}) belong to the category of the ``beyond Horndeski'' theories, which correspond to $m_4^2\neq {\tilde m}_4^2$ in action (\ref{action01}).
The quadratic action (\ref{QTcT002}) can be related to (\ref{tensor-action}) when $c_T=1$ and $Q_T=1$ by a disformal transformation $\hat{g}_{\mu\nu} \rightarrow \hat{c}_{T}^{-1}\left[\hat{g}_{\mu \nu}+\left(1-\hat{c}_{T}^{2}\right) \hat{n}_{\mu} \hat{n}_{\nu}\right]$, which indicates that $\hat{\tau}$ and $\hat{a}$ satisfy
\be
d\tau=\hat{c}_T d \hat{\tau}\,,\qquad a = \hat{c}_T^{-1/2} \hat{a}\,, \label{eq:atau01}
\ee
see \cite{Creminelli:2014wna} for details. Therefore, we find
\be
{\cal H}(\tau) = {1\over \hat{c}_T(\hat{\tau})}\[ \hat{\cal H}(\hat{\tau}) - {1\over 2}{d \hat{c}_T/d \hat{\tau} \over \hat{c}_T(\hat{\tau}) } \]\,. \label{eq:DisformalH}
\ee
In Sec. \ref{sec:cTsetup}, we have assumed that $\hat{c}_T(\hat{\tau})$ is constant during phase 1 and phase 3.
As a result, we can obtain $\lf({\bar{\cal H}_1/ \bar{\cal H}_2} \rt)^{3-2\nu_2}={\bar{\hat{c}}_{T,2}}/{\bar{\hat{c}}_{T,1}}$ and $H_{inf2}=\hat{H}_{inf2}\bar{\hat{c}}_{T,2}^{-1/2}$.
It is easy to identify that
\be
\hat{P}_{T,inf2}=P_{T,inf2}\,,
\ee
see Eq. (\ref{eq:PT}) and (\ref{eq:PTcTab-1}) for the definitions of $P_{T,inf2}$ and $\hat{P}_{T,inf2}$, respectively.
Therefore, the power spectra of the two scenarios coincide with each other at the leading order for the perturbation modes exited their horizon in phase 1 and phase 3.

Regarding the NEC-violating phase and the $\hat{c}_T$-diminishing phase, we can find $p=-{2\epsilon_2 \over 1+\epsilon_2}$ by using (\ref{eq:atau01}), which indicates that $\hat{\nu}_2=\nu_2$, provided there exists such a disformal transformation. Therefore, for perturbation modes that exited their horizon during the intermediate phase, their spectra indexes are same in these two scenarios at the leading order. However, there is a discrepancy in the coefficients of the leading order terms in Eq. (\ref{eq:PTNEC02}) and (\ref{eq:PTcT03}), which should be attributed to the discontinuity of $d\hat{c}_{T}(\hat{\tau})/d\hat{\tau}$. As can be seen from Eq. (\ref{eq:DisformalH}), even if $d\hat{c}_{T}(\hat{\tau})/d\hat{\tau}$ is continuous everywhere, the oversimplified parameterizations of $a$ (or ${\cal H}$) and $\hat{c}_T$ cannot result in the equivalence of these two scenarios around the matching time $\tau_1$ (or $\hat{\tau}_1$) and $\tau_2$ (or $\hat{\tau}_2$), unless ${d \hat{c}_T/d \hat{\tau} / \hat{c}_T}$ is negligible.
For this reason, we should focus on the comparison of spectra indices of these two scenarios, which are evaluated for ${\hat{\cal H}_1/\bar{\hat{c}}_{T,1} \ll k\ll \hat{\cal H}_2/\bar{\hat{c}}_{T,2} }$ so that they can hardly be affected by the details around the matching time.

%{\red That is to say, these two scenarios depicted by such oversimplified parameterizations can not be exactly related by a disformal transformation, especially around $\tau_1$ (or $\hat{\tau}_1$) and $\tau_2$ (or $\hat{\tau}_2$).}

%Furthermore,
In the NEC-violating scenario, $\nu_2=1/2+1/(1-\epsilon_2)$, which indicates that $1/2<\nu_2<3/2$ and
\be 0<n_T=3-2\nu_2<2\,. \label{NEC-nT}
\ee
However, in the $\hat{c}_T$-diminishing scenario, $\hat{\nu}_2=(3+2p)/(2+2p)$ with $p>0$, which results in $1<\hat{\nu}_2<3/2$ and
\be
0<\hat{n}_T=3-2\hat{\nu}_2<1\,, \label{hat-nT}
\ee
where the background evolution has been assumed as inflation. During inflation,
a blue-tilted power spectrum of primordial GWs with $0<n_T<1$ could be generated by either the NEC violation or the diminishment of the propagating speed of primordial GWs, while $n_T>1$ can only be attributed to the violation of NEC.
In fact, in the limit $p\rightarrow\infty$, the relation (\ref{eq:DisformalH}) will be invalid due to the divergence of ${d \hat{c}_T/d \hat{\tau} / \hat{c}_T}$, namely, the correspondence of the two scenarios will be broken.
This is also due to the fact that we have required an inflationary background in the $\hat{c}_T\neq1$ case.

\begin{figure}
	\centering
	\includegraphics[angle=0, width=0.9\textwidth]{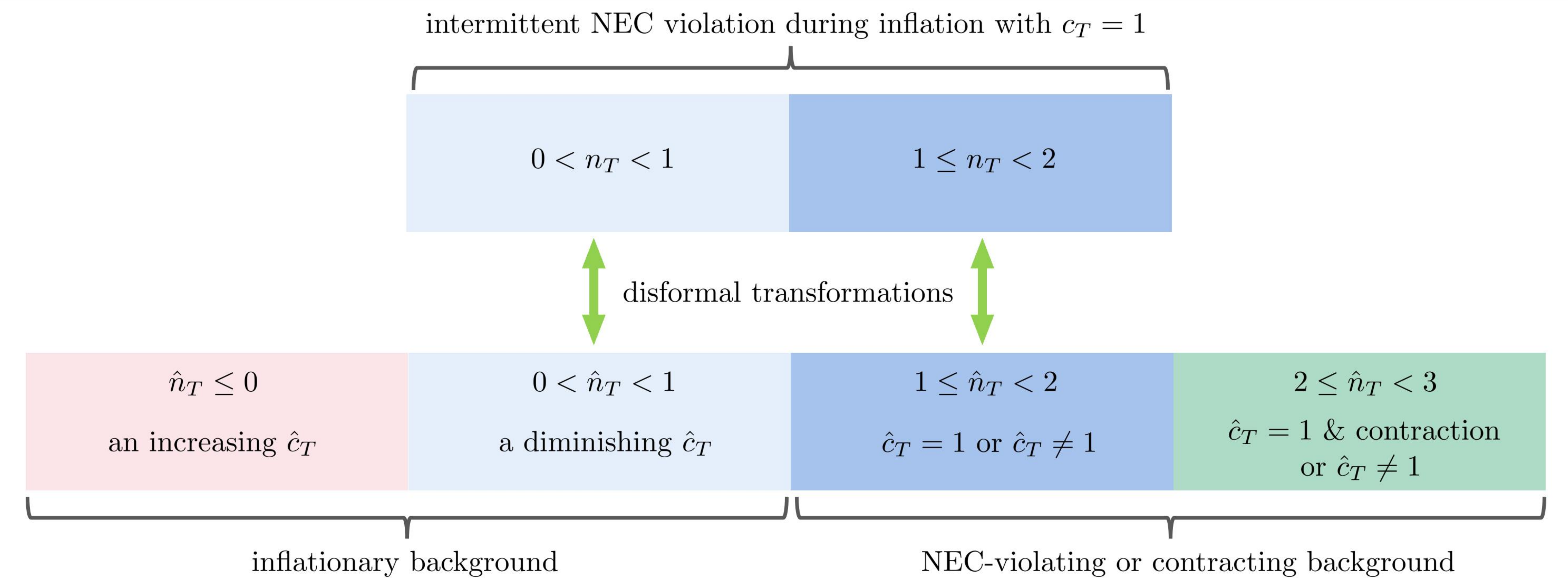}
	\caption{Relation and discrepancy between the NEC-violating scenario and the $c_T$-diminishing scenario. In the sense of the prediction of blue-tilted spectrum of primordial GWs with $0<n_T<1$, these two scenarios cannot be distinguished by observations of $n_T$ when there exists a disformal dual. However, the NEC-violating scenario with $c_T=1$ which predicts $1\leq n_T<2$ cannot be related to the $c_T\neq 1$ case by a disformal transformation, unless the background evolution is contracting or NEC-violating (namely, non-inflationary).}
	\label{fig:two-frames}
\end{figure}

The NEC-violating case with $\epsilon_2<-1$, which gives $1<n_T <2$, cannot be related to a varying $c_T$ with $p>-1$ on an inflationary background by a disformal transformation. To see this, we can assume that $\hat{a}_2(\hat{\tau})\sim \lf( \hat{\tau}_{R,j} - \hat{\tau} \rt)^{1\over \hat{\epsilon}_j-1}\sim \lf( \hat{\tau}_{R,j} - \hat{\tau} \rt)^{2\over 1+3\hat{w}_2}$ and $\hat{c}_{T,2}(\hat{\tau})\sim \lf( \hat{\tau}_{R,j} - \hat{\tau} \rt)^p$, where $\hat{w}_2=\hat{p}_2/\hat{\rho}_2$ is the equation of state parameter of phase 2. It can be checked that $p>-1$ is still required so that the tensor perturbation modes are able to exit their effective sound horizon during phase 2 ($\hat{\tau}_{2} < \hat{\tau} \leq \hat{\tau}_{3}$). Under the conditions $\epsilon_2<-1$ and $p>-1$, the disformal transformation (or the relation given by Eq. (\ref{eq:atau01})) actually gives $\hat{w}_2>-1/3$ or $\hat{w}_2<-5/3$, which indicates that the background evolution in the $\hat{c}_T\neq1$ frame should be either a contraction\footnote{A contracting Universe has to go through an NEC-violating bounce later to enter the Big Bang expansion.} or an NEC-violating expansion, namely non-inflationary, see Fig. \ref{fig:two-frames} for an illustration.
%Both of these two cases highlighted the significance of primordial NEC violation in generating a blue spectrum of GWs with $1<n_T <2$.
In other words, even if we give up the requirement of having an inflationary background in the $\hat{c}_T\neq1$ frame, we would conclude that an observation of $n_T>1$ would support the existence of a primordial NEC violation.

Since we have focused on the $\hat{c}_T$-diminishing scenario on the inflationary background in Sec. \ref{sec:cTdim}, the resulted range of $\hat{n}_T$ given by (\ref{hat-nT}) cannot cover that of the NEC-violating scenario (i.e., (\ref{NEC-nT})).
Therefore, in the sense of the prediction of blue-tilted primordial GWs, these two scenarios may be identified by observations of the spectra index when $n_T>1$, provided the background is assumed as inflation in the $\hat{c}_T\neq 1$ frame. Note that physics should not depend on the chosen frame. Our result does not mean the breakdown of the equivalence between the two scenarios which give $1\leq {n}_T<2$ in Fig. \ref{fig:two-frames}, but highlighted the significance of primordial NEC violation in generating GWs with $n_T>1$.

\begin{figure}[htbp]
    \subfigure[~~\label{figcTH-a}]{\includegraphics[width=.47\textwidth]{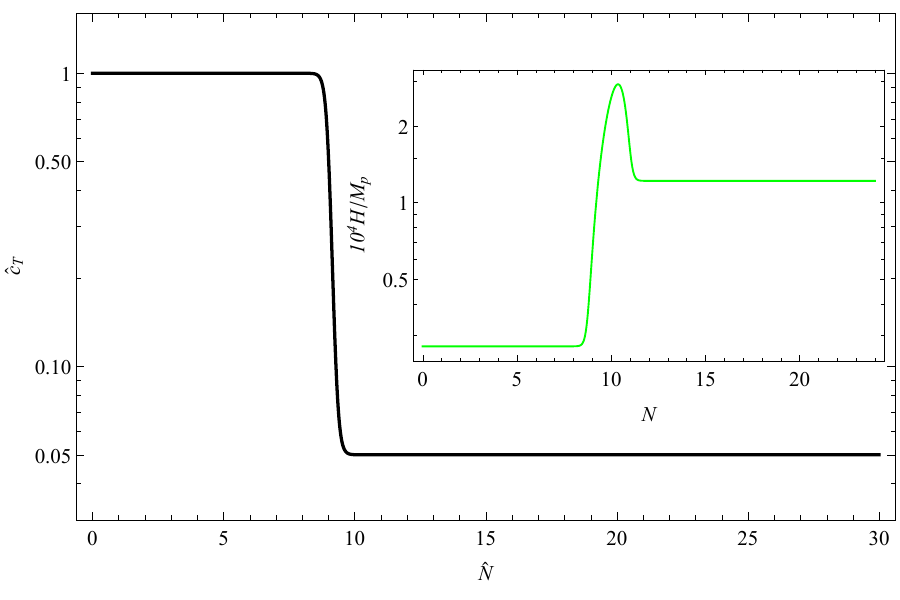} }
    \subfigure[~~\label{figcTH-b}]{\includegraphics[width=.47\textwidth]{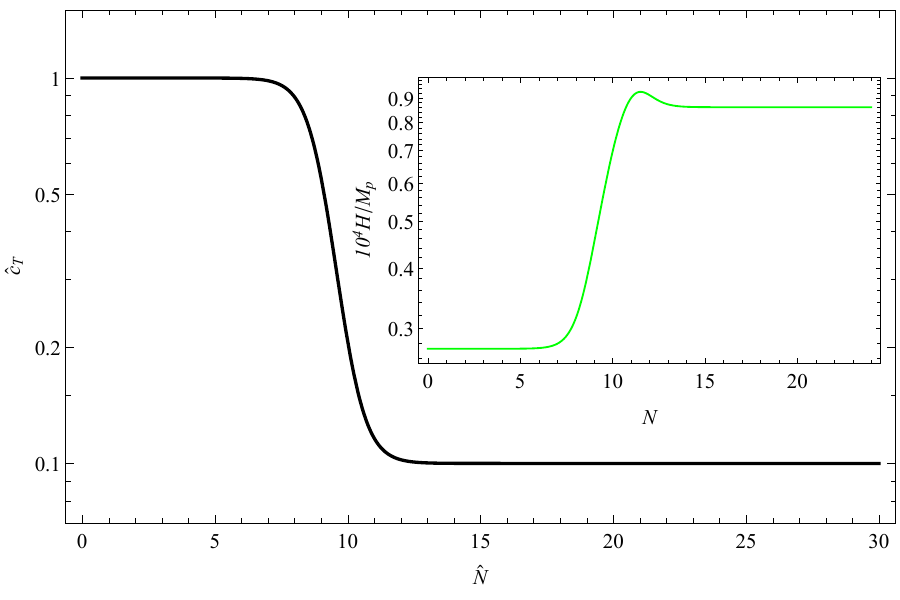} }
    \caption{The diminishing step-like $\hat{c}_T$ and the corresponding $H$, where the parameters $N$ and $\hat{N}$ satisfy $\Delta N=\Delta\ln a$ and $\Delta \hat{N}=\Delta\ln \hat{a}$, respectively. See Ref. \cite{Cai:2015yza} for more details about the numerical method.} \label{figcTH}
\end{figure}

\begin{figure}[htbp]
    \subfigure[~~]{\includegraphics[width=.47\textwidth]{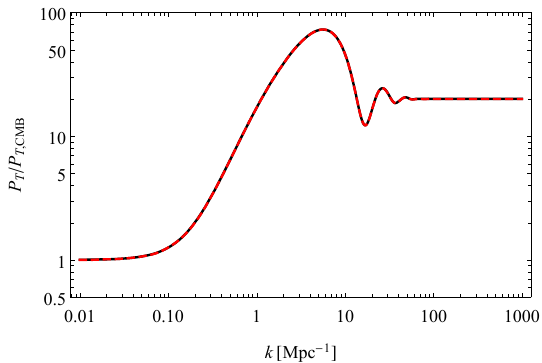} }
    \subfigure[~~\label{figPTframe-b}]{\includegraphics[width=.47\textwidth]{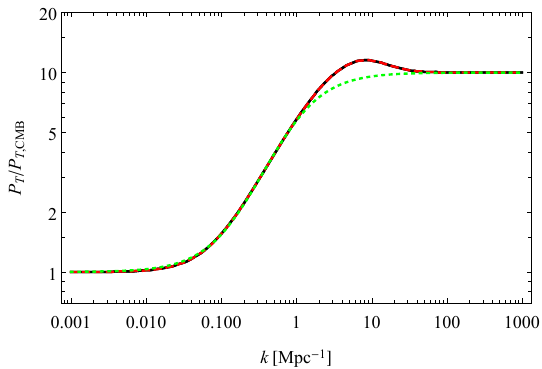} }
    \caption{Comparisons between the power spectra of primordial GWs in the NEC-violating scenario (red dashed curves) and the $\hat{c}_T$-diminishing scenario (black solid curves), where $P_{T,{\rm CMB}}$ is assumed to be the power spectrum of primordial GWs at the CMB frequency band. The green dotted curve is plotted by multiplying a factor 10 with the right hand side of Eq. (\ref{eq:para01}), where $\nu_2=0.88$, ${\cal H}_1=0.18\, {\rm Mpc}^{-1}$ and ${\cal H}_2=1.15\,{\rm Mpc}^{-1}$. Left: The power spectra induced by $\hat{c}_T$ and $H$ displayed in Fig. \ref{figcTH-a}. Right: The power spectra induced by $\hat{c}_T$ and $H$ displayed in Fig. \ref{figcTH-b}. See Ref. \cite{Cai:2015yza} for more details about the numerical method.} \label{figPTframe}
\end{figure}

In the context of $0<n_T<1$, we can numerically demonstrate the correspondence of the NEC-violating scenario and $\hat{c}_T$-diminishing scenario. For simplicity, we consider two examples of the step-like $\hat{c}_T$ and obtain the corresponding $H$ with Eqs. (\ref{eq:atau01}) and (\ref{eq:DisformalH}), which are displayed in Fig. \ref{figcTH}. Obviously, when ${d \hat{c}_T/d \hat{\tau} / \hat{c}_T}$ is significant, the corresponding $H$ can no longer be mimicked by the parameterization (\ref{a001}), see Fig. \ref{figcTH-a}.
The resulted power spectra in two scenarios exactly coincide with each other for smooth $\hat{c}_T$ regardless of the significance of ${d \hat{c}_T/d \hat{\tau} / \hat{c}_T}$, as displayed in Fig. \ref{figPTframe}, see also \cite{Cai:2015yza,Cai:2016ldn}. As a contrast, we also plotted the template (\ref{eq:para01}) in Fig. \ref{figPTframe-b}. Note that the numerical method used here follows that in \cite{Cai:2015yza}.

In order to be consistent with the current observations of LIGO and Virgo, the spectrum may decrease with respect to the frequency at much higher frequencies.
We can assume that $\hat{c}_T$ will regain the value of the speed of light or $H$ will go down to a smaller reasonable value before the end of inflation.


\section{Conclusion}\label{conclusion}

The history of the inflationary era could be much more complicated than expected \cite{Li:2019ipk,DAmico:2020euu,DAmico:2021zdd}.
The primordial GWs generated during inflation span a frequency band from $10^{-18}\,{\rm Hz}$ to $10^{10}\,{\rm Hz}$, hence could encode rich information about the history of the inflationary Universe.
Inflationary GWs with enhanced power spectra may be promising to explain the common-spectrum process reported recently by NANOGrav and PPTA.

Within the framework of EFT, we investigated an intermittent NEC violation as well as a diminishment of propagating speed of primordial GWs for generating an enhanced power spectrum of GW background during inflation. Both of the NEC violation and the time-dependent $\hat{c}_T$ can be realized stably in the ``beyond Horndeski'' theory.
We explicitly calculated the power spectra of primordial GWs for both the NEC-violating scenario and the ${c}_T$-diminishing scenario. Interestingly, the range of the spectra index in the NEC-violating scenario is $0<n_T<2$, while it is $0<n_T<1$ in the ${c}_T$-diminishing scenario. Therefore, these two scenarios can be identified by the spectra index when $n_T>1$, while they could be dual to each other by a disformal transformation of the metric in the sense of generating primordial GWs for the spectra index $n_T<1$.

Compared with previous works \cite{Cai:2020qpu,Creminelli:2014wna}, our work discussed explicitly the relation and discrepancy between two scenarios for different ranges of the spectra index $n_T$ of the primordial GW spectrum. We pointed out that the GW modes are not able to exit their effective sound horizon during the phase in which the propagating speed of GWs increases with time too fast (i.e., $p<-1$ in Eq. (\ref{eq:cT02})). As a result, a blue-tilted spectrum of GWs with $n_T>1$ cannot be attributed to a diminishing or increasing $c_T$ during inflation, unless the background evolution is replaced by non-inflationary ones, i.e., a contraction or NEC-violating expansion. Therefore, our work highlighted the significance of primordial NEC violation in generating a blue spectrum of GWs with $1<n_T <2$.

Our work reinforces the possibility that the primordial GWs could be non-negligible and essential for surveys of a stochastic GW background.
Future observational surveys of GWs at the frequency band of the CMB, PTA and interferometers could be combined to put constraints on the NEC-violating scenario as well as the ${c}_T$-varying scenario. The result of this paper can be generalized to the case where there are multiple violations of the NEC or nontrivial variations of $\hat{c}_T$.
Additionally, taking account of the enhancement of chirality of primordial GWs during inflation would also be interesting from both theoretical and observational standpoints \cite{Cai:2016ihp,Gao:2019liu,Obata:2016oym,Bartolo:2018elp,Nojiri:2019nar,Qiao:2019hkz,Mylova:2019jrj,Cai:2021uup,Kawai:2017kqt}.

%We explicitly calculated the power spectra of primordial GWs for a scenario in which the NEC-preserving slow-roll inflation is followed by a NEC-violating expanding phase and a later NEC-preserving slow-roll inflation with a larger Hubble parameter, while $c_T=1$ throughout. Templates of such spectra are provided. We also calculate the power spectra of primordial GWs for a scenario in which $c_T$ decreases with time during an intermediate phase of the slow-roll inflation, while the NEC is preserving throughout. The underlying relation between these two scenarios is discussed.

\acknowledgments

Y. C. would like to thank Zhibin Li and Ji Xu for helpful discussions.
The work of Y. C. was supported by the China Postdoctoral Science Foundation (Grant No. 2021M692942), the National Natural Science Foundation of China (Grant No. 11905224) and Zhengzhou University.
The work of Y.-S. P. was supported by the National Natural Science Foundation of China (Grants Nos. 12075246, 11690021). We acknowledge the use of the computing server {\it Arena317}@ZZU.

\appendix

%\begin{figure}[htbp]
%\centering %
%\includegraphics[scale=2,width=0.55\textwidth]{aH.eps}
%\caption{The evolutions of primordial perturbations. }\label{fig03}
%\end{figure}

%\section{Appendix}
%\label{Sec:Ap}

%%%%%%%%%%%%%%%%%%%%%%%%%%%%%%%%%%
% bibliography

%\bibliographystyle{plain}
\bibliography{Setting/ref}

\providecommand{\href}[2]{#2}\begingroup\raggedright\begin{thebibliography}{100}

\bibitem{LIGOScientific:2016aoc}
{\scshape LIGO Scientific, Virgo} collaboration, B.~P. Abbott et~al.,
  \emph{{Observation of Gravitational Waves from a Binary Black Hole Merger}},
  \href{https://doi.org/10.1103/PhysRevLett.116.061102}{\emph{Phys. Rev. Lett.}
  {\bfseries 116} (2016) 061102},
  [\href{https://arxiv.org/abs/1602.03837}{{\ttfamily 1602.03837}}].

\bibitem{LIGOScientific:2017vwq}
{\scshape LIGO Scientific, Virgo} collaboration, B.~P. Abbott et~al.,
  \emph{{GW170817: Observation of Gravitational Waves from a Binary Neutron
  Star Inspiral}},
  \href{https://doi.org/10.1103/PhysRevLett.119.161101}{\emph{Phys. Rev. Lett.}
  {\bfseries 119} (2017) 161101},
  [\href{https://arxiv.org/abs/1710.05832}{{\ttfamily 1710.05832}}].

\bibitem{Ellis:2020ena}
J.~Ellis and M.~Lewicki, \emph{{Cosmic String Interpretation of NANOGrav Pulsar
  Timing Data}},
  \href{https://doi.org/10.1103/PhysRevLett.126.041304}{\emph{Phys. Rev. Lett.}
  {\bfseries 126} (2021) 041304},
  [\href{https://arxiv.org/abs/2009.06555}{{\ttfamily 2009.06555}}].

\bibitem{Li:2020cjj}
H.-H. Li, G.~Ye and Y.-S. Piao, \emph{{Is the NANOGrav signal a hint of dS
  decay during inflation?}},
  \href{https://doi.org/10.1016/j.physletb.2021.136211}{\emph{Phys. Lett. B}
  {\bfseries 816} (2021) 136211},
  [\href{https://arxiv.org/abs/2009.14663}{{\ttfamily 2009.14663}}].

\bibitem{Addazi:2020zcj}
A.~Addazi, Y.-F. Cai, Q.~Gan, A.~Marciano and K.~Zeng, \emph{{NANOGrav results
  and dark first order phase transitions}},
  \href{https://doi.org/10.1007/s11433-021-1724-6}{\emph{Sci. China Phys. Mech.
  Astron.} {\bfseries 64} (2021) 290411},
  [\href{https://arxiv.org/abs/2009.10327}{{\ttfamily 2009.10327}}].

\bibitem{Tahara:2020fmn}
H.~W.~H. Tahara and T.~Kobayashi, \emph{{Nanohertz gravitational waves from a
  null-energy-condition violation in the early universe}},
  \href{https://doi.org/10.1103/PhysRevD.102.123533}{\emph{Phys. Rev. D}
  {\bfseries 102} (2020) 123533},
  [\href{https://arxiv.org/abs/2011.01605}{{\ttfamily 2011.01605}}].

\bibitem{Kuroyanagi:2020sfw}
S.~Kuroyanagi, T.~Takahashi and S.~Yokoyama, \emph{{Blue-tilted inflationary
  tensor spectrum and reheating in the light of NANOGrav results}},
  \href{https://doi.org/10.1088/1475-7516/2021/01/071}{\emph{JCAP} {\bfseries
  01} (2021) 071}, [\href{https://arxiv.org/abs/2011.03323}{{\ttfamily
  2011.03323}}].

\bibitem{Brandenburg:2021tmp}
A.~Brandenburg, E.~Clarke, Y.~He and T.~Kahniashvili, \emph{{Can we observe the
  QCD phase transition-generated gravitational waves through pulsar timing
  arrays?}}, \href{https://doi.org/10.1103/PhysRevD.104.043513}{\emph{Phys.
  Rev. D} {\bfseries 104} (2021) 043513},
  [\href{https://arxiv.org/abs/2102.12428}{{\ttfamily 2102.12428}}].

\bibitem{Yi:2021lxc}
Z.~Yi and Z.-H. Zhu, \emph{{NANOGrav signal and LIGO-Virgo Primordial Black
  Holes from Higgs inflation}},
  \href{https://arxiv.org/abs/2105.01943}{{\ttfamily 2105.01943}}.

\bibitem{Lewicki:2021xku}
M.~Lewicki, O.~Pujol\`as and V.~Vaskonen, \emph{{Escape from supercooling with
  or without bubbles: gravitational wave signatures}},
  \href{https://doi.org/10.1140/epjc/s10052-021-09669-6}{\emph{Eur. Phys. J. C}
  {\bfseries 81} (2021) 857},
  [\href{https://arxiv.org/abs/2106.09706}{{\ttfamily 2106.09706}}].

\bibitem{Bian:2020urb}
L.~Bian, R.-G. Cai, J.~Liu, X.-Y. Yang and R.~Zhou, \emph{{Evidence for
  different gravitational-wave sources in the NANOGrav dataset}},
  \href{https://doi.org/10.1103/PhysRevD.103.L081301}{\emph{Phys. Rev. D}
  {\bfseries 103} (2021) L081301},
  [\href{https://arxiv.org/abs/2009.13893}{{\ttfamily 2009.13893}}].

\bibitem{Sun:2021yra}
S.~Sun, X.-Y. Yang and Y.-L. Zhang, \emph{{Pulsar Timing Residual induced by
  Wideband Ultralight Dark Matter with Spin 0, 1, 2}},
  \href{https://arxiv.org/abs/2112.15593}{{\ttfamily 2112.15593}}.

\bibitem{Vagnozzi:2020gtf}
S.~Vagnozzi, \emph{{Implications of the NANOGrav results for inflation}},
  \href{https://doi.org/10.1093/mnrasl/slaa203}{\emph{Mon. Not. Roy. Astron.
  Soc.} {\bfseries 502} (2021) L11--L15},
  [\href{https://arxiv.org/abs/2009.13432}{{\ttfamily 2009.13432}}].

\bibitem{Samanta:2020cdk}
R.~Samanta and S.~Datta, \emph{{Gravitational wave complementarity and impact
  of NANOGrav data on gravitational leptogenesis}},
  \href{https://doi.org/10.1007/JHEP05(2021)211}{\emph{JHEP} {\bfseries 05}
  (2021) 211}, [\href{https://arxiv.org/abs/2009.13452}{{\ttfamily
  2009.13452}}].

\bibitem{Chen:2019xse}
Z.-C. Chen, C.~Yuan and Q.-G. Huang, \emph{{Pulsar Timing Array Constraints on
  Primordial Black Holes with NANOGrav 11-Year Dataset}},
  \href{https://doi.org/10.1103/PhysRevLett.124.251101}{\emph{Phys. Rev. Lett.}
  {\bfseries 124} (2020) 251101},
  [\href{https://arxiv.org/abs/1910.12239}{{\ttfamily 1910.12239}}].

\bibitem{Chen:2021wdo}
Z.-C. Chen, C.~Yuan and Q.-G. Huang, \emph{{Non-tensorial gravitational wave
  background in NANOGrav 12.5-year data set}},
  \href{https://doi.org/10.1007/s11433-021-1797-y}{\emph{Sci. China Phys. Mech.
  Astron.} {\bfseries 64} (2021) 120412},
  [\href{https://arxiv.org/abs/2101.06869}{{\ttfamily 2101.06869}}].

\bibitem{Wu:2021kmd}
Y.-M. Wu, Z.-C. Chen and Q.-G. Huang, \emph{{Constraining the Polarization of
  Gravitational Waves with the Parkes Pulsar Timing Array Second Data
  Release}},  \href{https://arxiv.org/abs/2108.10518}{{\ttfamily 2108.10518}}.

\bibitem{NANOGrav:2020bcs}
{\scshape NANOGrav} collaboration, Z.~Arzoumanian et~al., \emph{{The NANOGrav
  12.5 yr Data Set: Search for an Isotropic Stochastic Gravitational-wave
  Background}},
  \href{https://doi.org/10.3847/2041-8213/abd401}{\emph{Astrophys. J. Lett.}
  {\bfseries 905} (2020) L34},
  [\href{https://arxiv.org/abs/2009.04496}{{\ttfamily 2009.04496}}].

\bibitem{Goncharov:2021oub}
B.~Goncharov et~al., \emph{{On the evidence for a common-spectrum process in
  the search for the nanohertz gravitational-wave background with the Parkes
  Pulsar Timing Array}},  \href{https://arxiv.org/abs/2107.12112}{{\ttfamily
  2107.12112}}.

\bibitem{Starobinsky:1979ty}
A.~A. Starobinsky, \emph{{Spectrum of relict gravitational radiation and the
  early state of the universe}}, {\emph{JETP Lett.} {\bfseries 30} (1979)
  682--685}.

\bibitem{Rubakov:1982df}
V.~A. Rubakov, M.~V. Sazhin and A.~V. Veryaskin, \emph{{Graviton Creation in
  the Inflationary Universe and the Grand Unification Scale}},
  \href{https://doi.org/10.1016/0370-2693(82)90641-4}{\emph{Phys. Lett. B}
  {\bfseries 115} (1982) 189--192}.

\bibitem{Planck:2018vyg}
{\scshape Planck} collaboration, N.~Aghanim et~al., \emph{{Planck 2018 results.
  VI. Cosmological parameters}},
  \href{https://doi.org/10.1051/0004-6361/201833910}{\emph{Astron. Astrophys.}
  {\bfseries 641} (2020) A6},
  [\href{https://arxiv.org/abs/1807.06209}{{\ttfamily 1807.06209}}].

\bibitem{Cai:2020qpu}
Y.~Cai and Y.-S. Piao, \emph{{Intermittent null energy condition violations
  during inflation and primordial gravitational waves}},
  \href{https://doi.org/10.1103/PhysRevD.103.083521}{\emph{Phys. Rev. D}
  {\bfseries 103} (2021) 083521},
  [\href{https://arxiv.org/abs/2012.11304}{{\ttfamily 2012.11304}}].

\bibitem{Cai:2021yvq}
Y.-F. Cai, J.~Jiang, M.~Sasaki, V.~Vardanyan and Z.~Zhou, \emph{{Beating the
  Lyth Bound by Parametric Resonance during Inflation}},
  \href{https://doi.org/10.1103/PhysRevLett.127.251301}{\emph{Phys. Rev. Lett.}
  {\bfseries 127} (2021) 251301},
  [\href{https://arxiv.org/abs/2105.12554}{{\ttfamily 2105.12554}}].

\bibitem{Benetti:2021uea}
M.~Benetti, L.~L. Graef and S.~Vagnozzi, \emph{{Primordial gravitational waves
  from NANOGrav: a broken power-law approach}},
  \href{https://arxiv.org/abs/2111.04758}{{\ttfamily 2111.04758}}.

\bibitem{Piao:2003ty}
Y.-S. Piao and E.~Zhou, \emph{{Nearly scale invariant spectrum of adiabatic
  fluctuations may be from a very slowly expanding phase of the universe}},
  \href{https://doi.org/10.1103/PhysRevD.68.083515}{\emph{Phys. Rev. D}
  {\bfseries 68} (2003) 083515},
  [\href{https://arxiv.org/abs/hep-th/0308080}{{\ttfamily hep-th/0308080}}].

\bibitem{Piao:2004tq}
Y.-S. Piao and Y.-Z. Zhang, \emph{{Phantom inflation and primordial
  perturbation spectrum}},
  \href{https://doi.org/10.1103/PhysRevD.70.063513}{\emph{Phys. Rev. D}
  {\bfseries 70} (2004) 063513},
  [\href{https://arxiv.org/abs/astro-ph/0401231}{{\ttfamily
  astro-ph/0401231}}].

\bibitem{Baldi:2005gk}
M.~Baldi, F.~Finelli and S.~Matarrese, \emph{{Inflation with violation of the
  null energy condition}},
  \href{https://doi.org/10.1103/PhysRevD.72.083504}{\emph{Phys. Rev. D}
  {\bfseries 72} (2005) 083504},
  [\href{https://arxiv.org/abs/astro-ph/0505552}{{\ttfamily
  astro-ph/0505552}}].

\bibitem{Piao:2006jz}
Y.-S. Piao, \emph{{Gravitational wave background from phantom superinflation}},
  \href{https://doi.org/10.1103/PhysRevD.73.047302}{\emph{Phys. Rev. D}
  {\bfseries 73} (2006) 047302},
  [\href{https://arxiv.org/abs/gr-qc/0601115}{{\ttfamily gr-qc/0601115}}].

\bibitem{Li:2016awk}
H.-G. Li, Y.~Cai and Y.-S. Piao, \emph{{Towards the bounce inflationary
  gravitational wave}},
  \href{https://doi.org/10.1140/epjc/s10052-016-4554-2}{\emph{Eur. Phys. J. C}
  {\bfseries 76} (2016) 699},
  [\href{https://arxiv.org/abs/1605.09586}{{\ttfamily 1605.09586}}].

\bibitem{Creminelli:2010ba}
P.~Creminelli, A.~Nicolis and E.~Trincherini, \emph{{Galilean Genesis: An
  Alternative to inflation}},
  \href{https://doi.org/10.1088/1475-7516/2010/11/021}{\emph{JCAP} {\bfseries
  11} (2010) 021}, [\href{https://arxiv.org/abs/1007.0027}{{\ttfamily
  1007.0027}}].

\bibitem{Liu:2011ns}
Z.-G. Liu, J.~Zhang and Y.-S. Piao, \emph{{A Galileon Design of Slow
  Expansion}}, \href{https://doi.org/10.1103/PhysRevD.84.063508}{\emph{Phys.
  Rev. D} {\bfseries 84} (2011) 063508},
  [\href{https://arxiv.org/abs/1105.5713}{{\ttfamily 1105.5713}}].

\bibitem{Wang:2012bq}
Y.~Wang and R.~Brandenberger, \emph{{Scale-Invariant Fluctuations from Galilean
  Genesis}}, \href{https://doi.org/10.1088/1475-7516/2012/10/021}{\emph{JCAP}
  {\bfseries 10} (2012) 021},
  [\href{https://arxiv.org/abs/1206.4309}{{\ttfamily 1206.4309}}].

\bibitem{Liu:2012ww}
Z.-G. Liu and Y.-S. Piao, \emph{{A Galileon Design of Slow Expansion: Emergent
  universe}}, \href{https://doi.org/10.1016/j.physletb.2012.11.068}{\emph{Phys.
  Lett. B} {\bfseries 718} (2013) 734--739},
  [\href{https://arxiv.org/abs/1207.2568}{{\ttfamily 1207.2568}}].

\bibitem{Creminelli:2012my}
P.~Creminelli, K.~Hinterbichler, J.~Khoury, A.~Nicolis and E.~Trincherini,
  \emph{{Subluminal Galilean Genesis}},
  \href{https://doi.org/10.1007/JHEP02(2013)006}{\emph{JHEP} {\bfseries 02}
  (2013) 006}, [\href{https://arxiv.org/abs/1209.3768}{{\ttfamily 1209.3768}}].

\bibitem{Hinterbichler:2012fr}
K.~Hinterbichler, A.~Joyce, J.~Khoury and G.~E. Miller, \emph{{DBI Realizations
  of the Pseudo-Conformal Universe and Galilean Genesis Scenarios}},
  \href{https://doi.org/10.1088/1475-7516/2012/12/030}{\emph{JCAP} {\bfseries
  12} (2012) 030}, [\href{https://arxiv.org/abs/1209.5742}{{\ttfamily
  1209.5742}}].

\bibitem{Hinterbichler:2012yn}
K.~Hinterbichler, A.~Joyce, J.~Khoury and G.~E. Miller,
  \emph{{Dirac-Born-Infeld Genesis: An Improved Violation of the Null Energy
  Condition}},
  \href{https://doi.org/10.1103/PhysRevLett.110.241303}{\emph{Phys. Rev. Lett.}
  {\bfseries 110} (2013) 241303},
  [\href{https://arxiv.org/abs/1212.3607}{{\ttfamily 1212.3607}}].

\bibitem{Liu:2014tda}
Z.-G. Liu, H.~Li and Y.-S. Piao, \emph{{Preinflationary genesis with CMB B-mode
  polarization}}, \href{https://doi.org/10.1103/PhysRevD.90.083521}{\emph{Phys.
  Rev. D} {\bfseries 90} (2014) 083521},
  [\href{https://arxiv.org/abs/1405.1188}{{\ttfamily 1405.1188}}].

\bibitem{Pirtskhalava:2014esa}
D.~Pirtskhalava, L.~Santoni, E.~Trincherini and P.~Uttayarat, \emph{{Inflation
  from Minkowski Space}},
  \href{https://doi.org/10.1007/JHEP12(2014)151}{\emph{JHEP} {\bfseries 12}
  (2014) 151}, [\href{https://arxiv.org/abs/1410.0882}{{\ttfamily 1410.0882}}].

\bibitem{Nishi:2015pta}
S.~Nishi and T.~Kobayashi, \emph{{Generalized Galilean Genesis}},
  \href{https://doi.org/10.1088/1475-7516/2015/03/057}{\emph{JCAP} {\bfseries
  03} (2015) 057}, [\href{https://arxiv.org/abs/1501.02553}{{\ttfamily
  1501.02553}}].

\bibitem{Kobayashi:2015gga}
T.~Kobayashi, M.~Yamaguchi and J.~Yokoyama, \emph{{Galilean Creation of the
  Inflationary Universe}},
  \href{https://doi.org/10.1088/1475-7516/2015/07/017}{\emph{JCAP} {\bfseries
  07} (2015) 017}, [\href{https://arxiv.org/abs/1504.05710}{{\ttfamily
  1504.05710}}].

\bibitem{Cai:2016gjd}
Y.~Cai and Y.-S. Piao, \emph{{The slow expansion with nonminimal derivative
  coupling and its conformal dual}},
  \href{https://doi.org/10.1007/JHEP03(2016)134}{\emph{JHEP} {\bfseries 03}
  (2016) 134}, [\href{https://arxiv.org/abs/1601.07031}{{\ttfamily
  1601.07031}}].

\bibitem{Nishi:2016ljg}
S.~Nishi and T.~Kobayashi, \emph{{Scale-invariant perturbations from
  null-energy-condition violation: A new variant of Galilean genesis}},
  \href{https://doi.org/10.1103/PhysRevD.95.064001}{\emph{Phys. Rev. D}
  {\bfseries 95} (2017) 064001},
  [\href{https://arxiv.org/abs/1611.01906}{{\ttfamily 1611.01906}}].

\bibitem{Ageeva:2018lko}
Y.~A. Ageeva, O.~A. Evseev, O.~I. Melichev and V.~A. Rubakov, \emph{{Horndeski
  Genesis: strong coupling and absence thereof}},
  \href{https://doi.org/10.1051/epjconf/201819107010}{\emph{EPJ Web Conf.}
  {\bfseries 191} (2018) 07010},
  [\href{https://arxiv.org/abs/1810.00465}{{\ttfamily 1810.00465}}].

\bibitem{Mironov:2019qjt}
S.~Mironov, V.~Rubakov and V.~Volkova, \emph{{Genesis with general relativity
  asymptotics in beyond Horndeski theory}},
  \href{https://doi.org/10.1103/PhysRevD.100.083521}{\emph{Phys. Rev. D}
  {\bfseries 100} (2019) 083521},
  [\href{https://arxiv.org/abs/1905.06249}{{\ttfamily 1905.06249}}].

\bibitem{Ageeva:2020gti}
Y.~Ageeva, O.~Evseev, O.~Melichev and V.~Rubakov, \emph{{Toward evading the
  strong coupling problem in Horndeski genesis}},
  \href{https://doi.org/10.1103/PhysRevD.102.023519}{\emph{Phys. Rev. D}
  {\bfseries 102} (2020) 023519},
  [\href{https://arxiv.org/abs/2003.01202}{{\ttfamily 2003.01202}}].

\bibitem{Ilyas:2020zcb}
A.~Ilyas, M.~Zhu, Y.~Zheng and Y.-F. Cai, \emph{{Emergent Universe and Genesis
  from the DHOST Cosmology}},
  \href{https://arxiv.org/abs/2009.10351}{{\ttfamily 2009.10351}}.

\bibitem{Zhu:2021ggm}
M.~Zhu and Y.~Zheng, \emph{{Improved DHOST Genesis}},
  \href{https://doi.org/10.1007/JHEP11(2021)163}{\emph{JHEP} {\bfseries 11}
  (2021) 163}, [\href{https://arxiv.org/abs/2109.05277}{{\ttfamily
  2109.05277}}].

\bibitem{Cai:2016thi}
Y.~Cai, Y.~Wan, H.-G. Li, T.~Qiu and Y.-S. Piao, \emph{{The Effective Field
  Theory of nonsingular cosmology}},
  \href{https://doi.org/10.1007/JHEP01(2017)090}{\emph{JHEP} {\bfseries 01}
  (2017) 090}, [\href{https://arxiv.org/abs/1610.03400}{{\ttfamily
  1610.03400}}].

\bibitem{Creminelli:2016zwa}
P.~Creminelli, D.~Pirtskhalava, L.~Santoni and E.~Trincherini, \emph{{Stability
  of Geodesically Complete Cosmologies}},
  \href{https://doi.org/10.1088/1475-7516/2016/11/047}{\emph{JCAP} {\bfseries
  11} (2016) 047}, [\href{https://arxiv.org/abs/1610.04207}{{\ttfamily
  1610.04207}}].

\bibitem{Cai:2017tku}
Y.~Cai, H.-G. Li, T.~Qiu and Y.-S. Piao, \emph{{The Effective Field Theory of
  nonsingular cosmology: II}},
  \href{https://doi.org/10.1140/epjc/s10052-017-4938-y}{\emph{Eur. Phys. J. C}
  {\bfseries 77} (2017) 369},
  [\href{https://arxiv.org/abs/1701.04330}{{\ttfamily 1701.04330}}].

\bibitem{Cai:2017dyi}
Y.~Cai and Y.-S. Piao, \emph{{A covariant Lagrangian for stable nonsingular
  bounce}}, \href{https://doi.org/10.1007/JHEP09(2017)027}{\emph{JHEP}
  {\bfseries 09} (2017) 027},
  [\href{https://arxiv.org/abs/1705.03401}{{\ttfamily 1705.03401}}].

\bibitem{LIGOScientific:2018dkp}
{\scshape LIGO Scientific, Virgo} collaboration, B.~P. Abbott et~al.,
  \emph{{Tests of General Relativity with GW170817}},
  \href{https://doi.org/10.1103/PhysRevLett.123.011102}{\emph{Phys. Rev. Lett.}
  {\bfseries 123} (2019) 011102},
  [\href{https://arxiv.org/abs/1811.00364}{{\ttfamily 1811.00364}}].

\bibitem{Creminelli:2017sry}
P.~Creminelli and F.~Vernizzi, \emph{{Dark Energy after GW170817 and
  GRB170817A}},
  \href{https://doi.org/10.1103/PhysRevLett.119.251302}{\emph{Phys. Rev. Lett.}
  {\bfseries 119} (2017) 251302},
  [\href{https://arxiv.org/abs/1710.05877}{{\ttfamily 1710.05877}}].

\bibitem{Sakstein:2017xjx}
J.~Sakstein and B.~Jain, \emph{{Implications of the Neutron Star Merger
  GW170817 for Cosmological Scalar-Tensor Theories}},
  \href{https://doi.org/10.1103/PhysRevLett.119.251303}{\emph{Phys. Rev. Lett.}
  {\bfseries 119} (2017) 251303},
  [\href{https://arxiv.org/abs/1710.05893}{{\ttfamily 1710.05893}}].

\bibitem{Baker:2017hug}
T.~Baker, E.~Bellini, P.~G. Ferreira, M.~Lagos, J.~Noller and I.~Sawicki,
  \emph{{Strong constraints on cosmological gravity from GW170817 and GRB
  170817A}}, \href{https://doi.org/10.1103/PhysRevLett.119.251301}{\emph{Phys.
  Rev. Lett.} {\bfseries 119} (2017) 251301},
  [\href{https://arxiv.org/abs/1710.06394}{{\ttfamily 1710.06394}}].

\bibitem{Langlois:2017dyl}
D.~Langlois, R.~Saito, D.~Yamauchi and K.~Noui, \emph{{Scalar-tensor theories
  and modified gravity in the wake of GW170817}},
  \href{https://doi.org/10.1103/PhysRevD.97.061501}{\emph{Phys. Rev. D}
  {\bfseries 97} (2018) 061501},
  [\href{https://arxiv.org/abs/1711.07403}{{\ttfamily 1711.07403}}].

\bibitem{deRham:2018red}
C.~de~Rham and S.~Melville, \emph{{Gravitational Rainbows: LIGO and Dark Energy
  at its Cutoff}},
  \href{https://doi.org/10.1103/PhysRevLett.121.221101}{\emph{Phys. Rev. Lett.}
  {\bfseries 121} (2018) 221101},
  [\href{https://arxiv.org/abs/1806.09417}{{\ttfamily 1806.09417}}].

\bibitem{deRham:2019ctd}
C.~de~Rham and A.~J. Tolley, \emph{{Speed of gravity}},
  \href{https://doi.org/10.1103/PhysRevD.101.063518}{\emph{Phys. Rev. D}
  {\bfseries 101} (2020) 063518},
  [\href{https://arxiv.org/abs/1909.00881}{{\ttfamily 1909.00881}}].

\bibitem{deRham:2020zyh}
C.~de~Rham and A.~J. Tolley, \emph{{Causality in curved spacetimes: The speed
  of light and gravity}},
  \href{https://doi.org/10.1103/PhysRevD.102.084048}{\emph{Phys. Rev. D}
  {\bfseries 102} (2020) 084048},
  [\href{https://arxiv.org/abs/2007.01847}{{\ttfamily 2007.01847}}].

\bibitem{Tian:2019vkc}
S.~X. Tian and Z.-H. Zhu, \emph{{Quantization of the Nonstandard Propagating
  Gravitational Waves in the Cosmological Background}},
  \href{https://doi.org/10.1016/j.dark.2019.100418}{\emph{Phys. Dark Univ.}
  {\bfseries 27} (2020) 100418},
  [\href{https://arxiv.org/abs/1911.10902}{{\ttfamily 1911.10902}}].

\bibitem{Cai:2015dta}
Y.~Cai, Y.-T. Wang and Y.-S. Piao, \emph{{Oscillating modulation to B-mode
  polarization from varying propagating speed of primordial gravitational
  waves}}, \href{https://doi.org/10.1103/PhysRevD.91.103001}{\emph{Phys. Rev.
  D} {\bfseries 91} (2015) 103001},
  [\href{https://arxiv.org/abs/1501.06345}{{\ttfamily 1501.06345}}].

\bibitem{Cai:2015ipa}
Y.~Cai, Y.-T. Wang and Y.-S. Piao, \emph{{Oscillation in power spectrum of
  primordial gravitational wave as a signature of higher-order stringy
  corrections}}, \href{https://doi.org/10.1007/JHEP02(2016)059}{\emph{JHEP}
  {\bfseries 02} (2016) 059},
  [\href{https://arxiv.org/abs/1508.07114}{{\ttfamily 1508.07114}}].

\bibitem{Giare:2020vss}
W.~Giar\`e and F.~Renzi, \emph{{Propagating speed of primordial gravitational
  waves}}, \href{https://doi.org/10.1103/PhysRevD.102.083530}{\emph{Phys. Rev.
  D} {\bfseries 102} (2020) 083530},
  [\href{https://arxiv.org/abs/2007.04256}{{\ttfamily 2007.04256}}].

\bibitem{Bernal:2020ywq}
N.~Bernal, A.~Ghoshal, F.~Hajkarim and G.~Lambiase, \emph{{Primordial
  Gravitational Wave Signals in Modified Cosmologies}},
  \href{https://doi.org/10.1088/1475-7516/2020/11/051}{\emph{JCAP} {\bfseries
  11} (2020) 051}, [\href{https://arxiv.org/abs/2008.04959}{{\ttfamily
  2008.04959}}].

\bibitem{Cai:2020ovp}
Y.-F. Cai, C.~Lin, B.~Wang and S.-F. Yan, \emph{{Sound speed resonance of the
  stochastic gravitational wave background}},
  \href{https://doi.org/10.1103/PhysRevLett.126.071303}{\emph{Phys. Rev. Lett.}
  {\bfseries 126} (2021) 071303},
  [\href{https://arxiv.org/abs/2009.09833}{{\ttfamily 2009.09833}}].

\bibitem{Cai:2015yza}
Y.~Cai, Y.-T. Wang and Y.-S. Piao, \emph{{Is there an effect of a nontrivial
  $c_T$ during inflation?}},
  \href{https://doi.org/10.1103/PhysRevD.93.063005}{\emph{Phys. Rev. D}
  {\bfseries 93} (2016) 063005},
  [\href{https://arxiv.org/abs/1510.08716}{{\ttfamily 1510.08716}}].

\bibitem{Cai:2016ldn}
Y.~Cai, Y.-T. Wang and Y.-S. Piao, \emph{{Propagating speed of primordial
  gravitational waves and inflation}},
  \href{https://doi.org/10.1103/PhysRevD.94.043002}{\emph{Phys. Rev. D}
  {\bfseries 94} (2016) 043002},
  [\href{https://arxiv.org/abs/1602.05431}{{\ttfamily 1602.05431}}].

\bibitem{Giovannini:2018zbf}
M.~Giovannini, \emph{{The propagating speed of relic gravitational waves and
  their refractive index during inflation}},
  \href{https://doi.org/10.1140/epjc/s10052-018-5931-9}{\emph{Eur. Phys. J. C}
  {\bfseries 78} (2018) 442},
  [\href{https://arxiv.org/abs/1803.05203}{{\ttfamily 1803.05203}}].

\bibitem{Giovannini:2018nkt}
M.~Giovannini, \emph{{Blue and violet graviton spectra from a dynamical
  refractive index}},
  \href{https://doi.org/10.1016/j.physletb.2018.12.068}{\emph{Phys. Lett. B}
  {\bfseries 789} (2019) 502--507},
  [\href{https://arxiv.org/abs/1805.08142}{{\ttfamily 1805.08142}}].

\bibitem{Mishima:2019vlh}
Y.~Mishima and T.~Kobayashi, \emph{{Revisiting slow-roll dynamics and the
  tensor tilt in general single-field inflation}},
  \href{https://doi.org/10.1103/PhysRevD.101.043536}{\emph{Phys. Rev. D}
  {\bfseries 101} (2020) 043536},
  [\href{https://arxiv.org/abs/1911.02143}{{\ttfamily 1911.02143}}].

\bibitem{Capurri:2020qgz}
G.~Capurri, N.~Bartolo, D.~Maino and S.~Matarrese, \emph{{Let Effective Field
  Theory of Inflation flow: stochastic generation of models with red/blue
  tensor tilt}},
  \href{https://doi.org/10.1088/1475-7516/2020/11/037}{\emph{JCAP} {\bfseries
  11} (2020) 037}, [\href{https://arxiv.org/abs/2006.10781}{{\ttfamily
  2006.10781}}].

\bibitem{Giovannini:2021uvh}
M.~Giovannini, \emph{{The refractive index of the relic gravitons and the nHz
  band}},  \href{https://arxiv.org/abs/2112.10564}{{\ttfamily 2112.10564}}.

\bibitem{Horndeski:1974wa}
G.~W. Horndeski, \emph{{Second-order scalar-tensor field equations in a
  four-dimensional space}},
  \href{https://doi.org/10.1007/BF01807638}{\emph{Int. J. Theor. Phys.}
  {\bfseries 10} (1974) 363--384}.

\bibitem{Deffayet:2011gz}
C.~Deffayet, X.~Gao, D.~A. Steer and G.~Zahariade, \emph{{From k-essence to
  generalised Galileons}},
  \href{https://doi.org/10.1103/PhysRevD.84.064039}{\emph{Phys. Rev. D}
  {\bfseries 84} (2011) 064039},
  [\href{https://arxiv.org/abs/1103.3260}{{\ttfamily 1103.3260}}].

\bibitem{Kobayashi:2011nu}
T.~Kobayashi, M.~Yamaguchi and J.~Yokoyama, \emph{{Generalized G-inflation:
  Inflation with the most general second-order field equations}},
  \href{https://doi.org/10.1143/PTP.126.511}{\emph{Prog. Theor. Phys.}
  {\bfseries 126} (2011) 511--529},
  [\href{https://arxiv.org/abs/1105.5723}{{\ttfamily 1105.5723}}].

\bibitem{Gleyzes:2014dya}
J.~Gleyzes, D.~Langlois, F.~Piazza and F.~Vernizzi, \emph{{Healthy theories
  beyond Horndeski}},
  \href{https://doi.org/10.1103/PhysRevLett.114.211101}{\emph{Phys. Rev. Lett.}
  {\bfseries 114} (2015) 211101},
  [\href{https://arxiv.org/abs/1404.6495}{{\ttfamily 1404.6495}}].

\bibitem{Langlois:2015cwa}
D.~Langlois and K.~Noui, \emph{{Degenerate higher derivative theories beyond
  Horndeski: evading the Ostrogradski instability}},
  \href{https://doi.org/10.1088/1475-7516/2016/02/034}{\emph{JCAP} {\bfseries
  02} (2016) 034}, [\href{https://arxiv.org/abs/1510.06930}{{\ttfamily
  1510.06930}}].

\bibitem{Langlois:2017mxy}
D.~Langlois, M.~Mancarella, K.~Noui and F.~Vernizzi, \emph{{Effective
  Description of Higher-Order Scalar-Tensor Theories}},
  \href{https://doi.org/10.1088/1475-7516/2017/05/033}{\emph{JCAP} {\bfseries
  05} (2017) 033}, [\href{https://arxiv.org/abs/1703.03797}{{\ttfamily
  1703.03797}}].

\bibitem{Langlois:2017mdk}
D.~Langlois, \emph{{Degenerate Higher-Order Scalar-Tensor (DHOST) theories}},
  in \emph{{52nd Rencontres de Moriond on Gravitation}}, pp.~221--228, 2017,
  \href{https://arxiv.org/abs/1707.03625}{{\ttfamily 1707.03625}}.

\bibitem{Cheung:2007st}
C.~Cheung, P.~Creminelli, A.~L. Fitzpatrick, J.~Kaplan and L.~Senatore,
  \emph{{The Effective Field Theory of Inflation}},
  \href{https://doi.org/10.1088/1126-6708/2008/03/014}{\emph{JHEP} {\bfseries
  03} (2008) 014}, [\href{https://arxiv.org/abs/0709.0293}{{\ttfamily
  0709.0293}}].

\bibitem{Gubitosi:2012hu}
G.~Gubitosi, F.~Piazza and F.~Vernizzi, \emph{{The Effective Field Theory of
  Dark Energy}},
  \href{https://doi.org/10.1088/1475-7516/2013/02/032}{\emph{JCAP} {\bfseries
  02} (2013) 032}, [\href{https://arxiv.org/abs/1210.0201}{{\ttfamily
  1210.0201}}].

\bibitem{Gleyzes:2013ooa}
J.~Gleyzes, D.~Langlois, F.~Piazza and F.~Vernizzi, \emph{{Essential Building
  Blocks of Dark Energy}},
  \href{https://doi.org/10.1088/1475-7516/2013/08/025}{\emph{JCAP} {\bfseries
  08} (2013) 025}, [\href{https://arxiv.org/abs/1304.4840}{{\ttfamily
  1304.4840}}].

\bibitem{Piazza:2013coa}
F.~Piazza and F.~Vernizzi, \emph{{Effective Field Theory of Cosmological
  Perturbations}},
  \href{https://doi.org/10.1088/0264-9381/30/21/214007}{\emph{Class. Quant.
  Grav.} {\bfseries 30} (2013) 214007},
  [\href{https://arxiv.org/abs/1307.4350}{{\ttfamily 1307.4350}}].

\bibitem{Turner:1993vb}
M.~S. Turner, M.~J. White and J.~E. Lidsey, \emph{{Tensor perturbations in
  inflationary models as a probe of cosmology}},
  \href{https://doi.org/10.1103/PhysRevD.48.4613}{\emph{Phys. Rev. D}
  {\bfseries 48} (1993) 4613--4622},
  [\href{https://arxiv.org/abs/astro-ph/9306029}{{\ttfamily
  astro-ph/9306029}}].

\bibitem{Boyle:2005se}
L.~A. Boyle and P.~J. Steinhardt, \emph{{Probing the early universe with
  inflationary gravitational waves}},
  \href{https://doi.org/10.1103/PhysRevD.77.063504}{\emph{Phys. Rev. D}
  {\bfseries 77} (2008) 063504},
  [\href{https://arxiv.org/abs/astro-ph/0512014}{{\ttfamily
  astro-ph/0512014}}].

\bibitem{Zhao:2006mm}
W.~Zhao and Y.~Zhang, \emph{{Relic gravitational waves and their detection}},
  \href{https://doi.org/10.1103/PhysRevD.74.043503}{\emph{Phys. Rev. D}
  {\bfseries 74} (2006) 043503},
  [\href{https://arxiv.org/abs/astro-ph/0604458}{{\ttfamily
  astro-ph/0604458}}].

\bibitem{Kuroyanagi:2014nba}
S.~Kuroyanagi, T.~Takahashi and S.~Yokoyama, \emph{{Blue-tilted Tensor Spectrum
  and Thermal History of the Universe}},
  \href{https://doi.org/10.1088/1475-7516/2015/02/003}{\emph{JCAP} {\bfseries
  02} (2015) 003}, [\href{https://arxiv.org/abs/1407.4785}{{\ttfamily
  1407.4785}}].

\bibitem{Liu:2015psa}
X.-J. Liu, W.~Zhao, Y.~Zhang and Z.-H. Zhu, \emph{{Detecting Relic
  Gravitational Waves by Pulsar Timing Arrays: Effects of Cosmic Phase
  Transitions and Relativistic Free-Streaming Gases}},
  \href{https://doi.org/10.1103/PhysRevD.93.024031}{\emph{Phys. Rev. D}
  {\bfseries 93} (2016) 024031},
  [\href{https://arxiv.org/abs/1509.03524}{{\ttfamily 1509.03524}}].

\bibitem{Kolevatov:2017voe}
R.~Kolevatov, S.~Mironov, N.~Sukhov and V.~Volkova, \emph{{Cosmological bounce
  and Genesis beyond Horndeski}},
  \href{https://doi.org/10.1088/1475-7516/2017/08/038}{\emph{JCAP} {\bfseries
  08} (2017) 038}, [\href{https://arxiv.org/abs/1705.06626}{{\ttfamily
  1705.06626}}].

\bibitem{Piao:2003zm}
Y.-S. Piao, B.~Feng and X.-m. Zhang, \emph{{Suppressing CMB quadrupole with a
  bounce from contracting phase to inflation}},
  \href{https://doi.org/10.1103/PhysRevD.69.103520}{\emph{Phys. Rev. D}
  {\bfseries 69} (2004) 103520},
  [\href{https://arxiv.org/abs/hep-th/0310206}{{\ttfamily hep-th/0310206}}].

\bibitem{Piao:2005ag}
Y.-S. Piao, \emph{{A Possible explanation to low CMB quadrupole}},
  \href{https://doi.org/10.1103/PhysRevD.71.087301}{\emph{Phys. Rev. D}
  {\bfseries 71} (2005) 087301},
  [\href{https://arxiv.org/abs/astro-ph/0502343}{{\ttfamily
  astro-ph/0502343}}].

\bibitem{Liu:2013kea}
Z.-G. Liu, Z.-K. Guo and Y.-S. Piao, \emph{{Obtaining the CMB anomalies with a
  bounce from the contracting phase to inflation}},
  \href{https://doi.org/10.1103/PhysRevD.88.063539}{\emph{Phys. Rev. D}
  {\bfseries 88} (2013) 063539},
  [\href{https://arxiv.org/abs/1304.6527}{{\ttfamily 1304.6527}}].

\bibitem{Qiu:2015nha}
T.~Qiu and Y.-T. Wang, \emph{{G-Bounce Inflation: Towards Nonsingular Inflation
  Cosmology with Galileon Field}},
  \href{https://doi.org/10.1007/JHEP04(2015)130}{\emph{JHEP} {\bfseries 04}
  (2015) 130}, [\href{https://arxiv.org/abs/1501.03568}{{\ttfamily
  1501.03568}}].

\bibitem{Cai:2017pga}
Y.~Cai, Y.-T. Wang, J.-Y. Zhao and Y.-S. Piao, \emph{{Primordial perturbations
  with pre-inflationary bounce}},
  \href{https://doi.org/10.1103/PhysRevD.97.103535}{\emph{Phys. Rev. D}
  {\bfseries 97} (2018) 103535},
  [\href{https://arxiv.org/abs/1709.07464}{{\ttfamily 1709.07464}}].

\bibitem{Cai:2019hge}
Y.~Cai and Y.-S. Piao, \emph{{Pre-inflation and trans-Planckian censorship}},
  \href{https://doi.org/10.1007/s11433-020-1573-5}{\emph{Sci. China Phys. Mech.
  Astron.} {\bfseries 63} (2020) 110411},
  [\href{https://arxiv.org/abs/1909.12719}{{\ttfamily 1909.12719}}].

\bibitem{Creminelli:2014wna}
P.~Creminelli, J.~Gleyzes, J.~Nore\~na and F.~Vernizzi, \emph{{Resilience of
  the standard predictions for primordial tensor modes}},
  \href{https://doi.org/10.1103/PhysRevLett.113.231301}{\emph{Phys. Rev. Lett.}
  {\bfseries 113} (2014) 231301},
  [\href{https://arxiv.org/abs/1407.8439}{{\ttfamily 1407.8439}}].

\bibitem{Ijjas:2015zma}
A.~Ijjas and P.~J. Steinhardt, \emph{{The anamorphic universe}},
  \href{https://doi.org/10.1088/1475-7516/2015/10/001}{\emph{JCAP} {\bfseries
  10} (2015) 001}, [\href{https://arxiv.org/abs/1507.03875}{{\ttfamily
  1507.03875}}].

\bibitem{Li:2019ipk}
H.-H. Li, G.~Ye, Y.~Cai and Y.-S. Piao, \emph{{Trans-Planckian censorship of
  multistage inflation and dark energy}},
  \href{https://doi.org/10.1103/PhysRevD.101.063527}{\emph{Phys. Rev. D}
  {\bfseries 101} (2020) 063527},
  [\href{https://arxiv.org/abs/1911.06148}{{\ttfamily 1911.06148}}].

\bibitem{DAmico:2020euu}
G.~D'Amico and N.~Kaloper, \emph{{Rollercoaster cosmology}},
  \href{https://doi.org/10.1088/1475-7516/2021/08/058}{\emph{JCAP} {\bfseries
  08} (2021) 058}, [\href{https://arxiv.org/abs/2011.09489}{{\ttfamily
  2011.09489}}].

\bibitem{DAmico:2021zdd}
G.~D'Amico, N.~Kaloper and A.~Westphal, \emph{{Very Hairy Inflation}},
  \href{https://arxiv.org/abs/2112.13861}{{\ttfamily 2112.13861}}.

\bibitem{Cai:2016ihp}
Y.~Cai, Y.-T. Wang and Y.-S. Piao, \emph{{Chirality oscillation of primordial
  gravitational waves during inflation}},
  \href{https://doi.org/10.1007/JHEP03(2017)024}{\emph{JHEP} {\bfseries 03}
  (2017) 024}, [\href{https://arxiv.org/abs/1608.06508}{{\ttfamily
  1608.06508}}].

\bibitem{Gao:2019liu}
X.~Gao and X.-Y. Hong, \emph{{Propagation of gravitational waves in a
  cosmological background}},
  \href{https://doi.org/10.1103/PhysRevD.101.064057}{\emph{Phys. Rev. D}
  {\bfseries 101} (2020) 064057},
  [\href{https://arxiv.org/abs/1906.07131}{{\ttfamily 1906.07131}}].

\bibitem{Obata:2016oym}
I.~Obata, \emph{{Chiral primordial blue tensor spectra from the axion-gauge
  couplings}}, \href{https://doi.org/10.1088/1475-7516/2017/06/050}{\emph{JCAP}
  {\bfseries 06} (2017) 050},
  [\href{https://arxiv.org/abs/1612.08817}{{\ttfamily 1612.08817}}].

\bibitem{Bartolo:2018elp}
N.~Bartolo, G.~Orlando and M.~Shiraishi, \emph{{Measuring chiral gravitational
  waves in Chern-Simons gravity with CMB bispectra}},
  \href{https://doi.org/10.1088/1475-7516/2019/01/050}{\emph{JCAP} {\bfseries
  01} (2019) 050}, [\href{https://arxiv.org/abs/1809.11170}{{\ttfamily
  1809.11170}}].

\bibitem{Nojiri:2019nar}
S.~Nojiri, S.~D. Odintsov, V.~K. Oikonomou and A.~A. Popov, \emph{{Propagation
  of Gravitational Waves in Chern-Simons Axion Einstein Gravity}},
  \href{https://doi.org/10.1103/PhysRevD.100.084009}{\emph{Phys. Rev. D}
  {\bfseries 100} (2019) 084009},
  [\href{https://arxiv.org/abs/1909.01324}{{\ttfamily 1909.01324}}].

\bibitem{Qiao:2019hkz}
J.~Qiao, T.~Zhu, W.~Zhao and A.~Wang, \emph{{Polarized primordial gravitational
  waves in the ghost-free parity-violating gravity}},
  \href{https://doi.org/10.1103/PhysRevD.101.043528}{\emph{Phys. Rev. D}
  {\bfseries 101} (2020) 043528},
  [\href{https://arxiv.org/abs/1911.01580}{{\ttfamily 1911.01580}}].

\bibitem{Mylova:2019jrj}
M.~Mylova, \emph{{Chiral primordial gravitational waves in extended theories of
  Scalar-Tensor gravity}},  \href{https://arxiv.org/abs/1912.00800}{{\ttfamily
  1912.00800}}.

\bibitem{Cai:2021uup}
R.-G. Cai, C.~Fu and W.-W. Yu, \emph{{Parity violation in stochastic
  gravitational wave background from inflation}},
  \href{https://arxiv.org/abs/2112.04794}{{\ttfamily 2112.04794}}.

\bibitem{Kawai:2017kqt}
S.~Kawai and J.~Kim, \emph{{Gauss\textendash{}Bonnet Chern\textendash{}Simons
  gravitational wave leptogenesis}},
  \href{https://doi.org/10.1016/j.physletb.2018.12.019}{\emph{Phys. Lett. B}
  {\bfseries 789} (2019) 145--149},
  [\href{https://arxiv.org/abs/1702.07689}{{\ttfamily 1702.07689}}].

\end{thebibliography}\endgroup
\bibliographystyle{Setting/jhep}

\end{document}